\documentclass[aps,pre,reprint]{revtex4-1}

\usepackage[utf8]{inputenc}
\usepackage{amsmath}
\usepackage{amsfonts}
\usepackage{amssymb}
\usepackage{amsthm}
\usepackage{amsopn}
\usepackage{upgreek}
\usepackage{bm}
\usepackage{epsfig}
\usepackage{units}
\usepackage{enumitem}

\setcitestyle{round,numbers,sort&compress}
\renewcommand{\eqref}[1]{Eq.~(\ref{#1})}
\newcommand{\eqsref}[1]{Eqs.~(\ref{#1})}

\newcommand{\figref}[1]{Figure~\ref{#1}}
\newcommand{\figsref}[1]{Figures~\ref{#1}}
\newcommand{\tabref}[1]{Table~\ref{#1}}
\newcommand{\tabsref}[1]{Tables~\ref{#1}}
\newcommand{\secref}[1]{Sec.~\ref{#1}}

\newcommand{\refcite}[1]{Ref.~\onlinecite{#1}}
\newcommand{\refscite}[1]{Refs.~\onlinecite{#1}}

\newcommand{\kB}{k_{\text{B}}}
\newcommand{\DSl}{\Delta S_{\text{l}}}
\newcommand{\kfold}{k_{\text{fold}}}

\newcommand{\pfold}{p_{\text{fold}}}

\newcommand{\addressharvard}{Department of Chemistry and Chemical Biology, Harvard University, 12 Oxford Street, Cambridge, MA 02138, USA}

\begin{document}

\title{Structure-based prediction of protein-folding transition paths}
\author{William M.~Jacobs}
\affiliation{\addressharvard}
\author{Eugene I.~Shakhnovich}
\affiliation{\addressharvard}
\date{\today}

\begin{abstract}
  We propose a general theory to describe the distribution of protein-folding transition paths.
  We show that transition paths follow a predictable sequence of high-free-energy transient states that are separated by free-energy barriers.
  Each transient state corresponds to the assembly of one or more discrete, cooperative units, which are determined directly from the native structure.
  We show that the transition state on a folding pathway is reached when a small number of critical contacts are formed between a specific set of substructures, after which folding proceeds downhill in free energy.
  This approach suggests a natural resolution for distinguishing parallel folding pathways and provides a simple means to predict the rate-limiting step in a folding reaction.
  Our theory identifies a common folding mechanism for proteins with diverse native structures and establishes general principles for the self-assembly of polymers with specific interactions.
\end{abstract}

\maketitle

\section{Introduction}

Protein folding has been described as both exceedingly complex and remarkably simple~\cite{sali1994how,shakhnovich1994proteins,onuchic2004theory,shakhnovich2006protein,daggett2003present,thirumalai2010theoretical}.
Although kinetic measurements are often consistent with simple two-state folding behavior~\cite{jackson1991folding}, experiments probing folding at higher resolution have provided evidence of considerable additional complexity~\cite{maity2005protein,englander2014nature}.
Direct observations of folding transition paths in both simulation~\cite{shaw2010atomic,lindorff2011fast} and experiment~\cite{chung2012single,hu2013stepwise,neupane2016direct}, including demonstrations that folding pathways can be redirected under various conditions~\cite{mickler2007revealing,jagannathan2012direct,guinn2015single}, can provide insight into these crucial yet fleeting events.
However, the factors that determine the distribution of folding transition paths and the detailed kinetics along these pathways remain poorly understood.

To address this question, we propose a general theory to predict the folding transition paths of globular proteins.
We adopt a simplified representation of a protein based on native contacts that are derived from a crystal structure~\cite{taketomi1975studies}.
Discrete `Ising-like' models~\cite{munoz1999simple,alm1999prediction,galzitskaya1999theoretical} have had great success in reproducing a wide variety of experimental measurements~\cite{kubelka2008chemical,henry2013comparing} without computationally expensive simulations.
However, due to the inherent combinatorial complexity of such models, previous studies have relied on the simplifying assumption that regions of native structure can only grow in one or two contiguous sequences.
This assumption is justified for very small proteins on the basis of helix--coil theory, but it limits the applicability of Ising-like models to proteins with relatively simple native-state topologies.
Here we take an alternative approach that enforces the intrinsic kinetic connectivity of the microstates and allows for a much larger space of physically realistic combinations of native contacts.
As a result, we are able to show that proteins fold by assembling discrete substructures via a small number of well-defined pathways.
In contrast to previous studies, our assumptions do not impose a specific mechanism of folding and are thus applicable to proteins with complex native-state topologies.

Our central finding is that folding can be described as a predictable sequence of transitions between discrete transient states.
First, we explain how kinetically distinct transient states can be predicted on the basis of a protein's native-state topology by developing a principle of substructure cooperativity.
We then show that the resulting network of transient states leads to a mechanistic description of protein-folding transition paths.
As a consequence, we are able to distinguish the small set of native contacts that are made precisely at the rate-limiting step from the many contacts that are formed earlier on a folding transition path.
As an example, we apply our theory to ubiquitin, a 76-residue $\alpha/\beta$ protein, for which detailed atomistic folding simulations and experimental characterizations are available.
We then show that our predictions are consistent with kinetic measurements on a large number of proteins.
Our results have implications both for understanding the folding transition paths of naturally occurring proteins at a detailed level and, more generally, for manipulating the self-assembly pathways of designed polymers with specific interactions.\vskip-1ex

\section{Theory}
\label{sec:theory}

In order for a protein to fold to a thermodynamically stable structure, the native state must be stabilized by a large energy gap relative to the many alternative configurations~\cite{sali1994how,shakhnovich1994proteins,onuchic2004theory,shakhnovich2006protein}.
Analysis of atomistic folding simulations provides strong evidence that the native contacts also play a central role in determining protein-folding transition paths~\cite{best2013native}.
Here we develop a native-centric, coarse-grained polymer model, where the pairwise contacts that define the completely folded state are associated with energetically favorable bonds.
We define native residue--residue interactions according to a fixed cutoff distance (4~\AA) between heavy atoms in a crystallographically determined native structure (\figsref{fig:graph}a and b).
We further restrict these interactions to residues that are more than one Kuhn length, taken here to be two residues, apart in the protein sequence.
This excludes native contacts that are typically not independent due to their close proximity and are likely to be present in the unfolded state.

The essential advantage of our theory is the identification of kinetically distinct transient states.
This aspect is crucial because it allows us to define a free-energy landscape that preserves the kinetic connectivity of microstates in the full combinatorial model.
Moreover, this reduction of complexity to a smaller number of coarse-grained configurations allows us to obtain a mechanistic description of protein-folding transition paths.
In the following sections, we outline the steps required for this approach.
First, we describe the statistical mechanics of the model and the choices of adjustable energetic parameters.
We then explain the physical justification for decomposing a protein into discrete, cooperative substructures, which contribute to the kinetically distinct states.
(Free-energy calculations and evidence from atomistic Molecular Dynamics simulations in support of our approach are presented in \secref{sec:results}.)
Lastly, we show how these coarse-grained states can be incorporated into a master-equation framework for predicting protein-folding transition paths.

\subsection*{Contact-graph model}

\begin{figure}
  \includegraphics{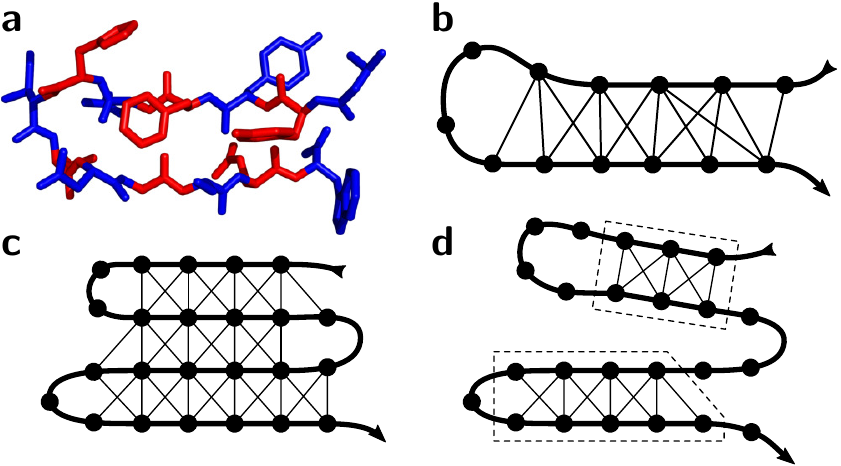}
  \vskip-1ex
  \caption{\textbf{Construction of the contact-graph model.}  (a) A portion of a $\beta$-hairpin, with sequential residues indicated by alternating colors.  We assume that the residues are segmented at the N--C$\alpha$ bond.  (b) An abstract graph representation of this structure, where vertices correspond to residues and edges to residue--residue contacts.  The polymer backbone is indicated by the heavy line.  (c) A schematic contact graph and (d) an allowed microstate, with independent structured regions indicated by dashes.  Within each structured region, all possible native contacts are formed.}
  \label{fig:graph}
\end{figure}

Microstates in this discrete model refer to coarse-grained representations of the polymer: each microstate comprises an ensemble of microscopic polymer configurations in which the residues make a specific combination of native contacts.
The microstate in which all specific contacts are formed corresponds to the completely folded configuration, while microstates with a subset of specific contacts are associated with partially folded configurations.
However, not all combinations of contacts correspond to physical configurations, since the conformational space of the polymer is restricted by steric constraints (the residues occupy finite volumes that cannot overlap) and the chain connectivity (sequential residues are covalently linked).
We therefore limit the set of allowed microstates to physically realistic configurations by imposing two rules.
First, we note that every microstate with a specific set of contacts can be decomposed into disconnected structured regions (\figref{fig:graph}c).
Within each structured region, it is reasonable to assume that the native contacts between interacting residues are geometrically correlated due to their close spatial proximity.
We therefore require that all possible native contacts be formed within each structured region (\figref{fig:graph}d).
Second, in order to define a self-consistent configurational entropy, we do not allow microstates with disordered loops of contact-forming residues (i.e., residues that make contacts in the native state) that are shorter than one Kuhn length (see SI text).

Because the microstates correspond to ensembles of constrained polymer configurations, each microstate $g$ is associated with a free energy, ${F(g)}$,
\begin{equation}
  \frac{F(g)}{\kB T} = \!\!\! \sum_{c \in \mathcal{C}(g)} \!\!\! \left\{ \!\! (N_c - 1) \frac{\mu}{\kB T} + \!\! \sum_{(u,v)} \!\! \frac{\bm{1}_{uv}^c \epsilon_{uv}^{\phantom{c}}}{\kB T} \! \right\} - \frac{\DSl(g)}{\kB},
  \label{eq:Fg}
\end{equation}
where $N_c$ is the number of residues in each structured region ${c \in \mathcal{C}(g)}$, $T$ is the absolute temperature and $\kB$ is the Boltzmann constant.
Within each structured region, we account for the loss of configurational entropy per ordered residue, $\mu/T$, and the energetic contributions, $\{\epsilon_{uv}\}$, of all native contacts.
(The notation $\bm{1}_{uv}^c$ indicates unity if a native contact is present between residues $u$ and $v$ in structured region $c$, and zero otherwise.)
The remaining entropic penalty, $\DSl$, accounts for closed loops of non-interacting residues.
Assuming Gaussian polymer statistics~\cite{vanderzande1998lattice} for sequences longer than one Kuhn length $b$, we sum the entropic penalties for all loops,
\begin{equation}
  \frac{\DSl(g)}{\kB} \equiv \!\!\sum_{l \in \mathcal{L}(g)} \!\!
  \begin{cases}
    -\frac{|l| \mu}{\kB T}
    & \!\!\!\text{if } |l| \le b, \\
    -\frac{b \mu}{\kB T} - \frac{d}{2} \! \left[ \ln \frac{|l|}{b} + \frac{r(l)^2}{b^2 |l|} \right]
    & \!\!\!\text{if } |l| > b,
  \end{cases}
  \label{eq:Sloop}
\end{equation}
where the sum runs over every loop in the microstate $g$, ${l \in \mathcal{L}(g)}$, ${|l|}$ is the number of non-interacting residues in the loop, $r(l)$ is the distance between the fixed ends of the loop, and ${d = 3}$ is the spatial dimension (see SI text).

In order to apply \eqsref{eq:Fg} and (\ref{eq:Sloop}), we must choose the parameters $\mu$ and $\{\epsilon_{uv}\}$.
On the basis of atomistic simulations~\cite{baxa2014loss}, we have chosen ${\mu = 2 \kB T}$; values between ${1.5 \kB T}$ and ${2.5 \kB T}$ give very similar results.
The energy of each bond is estimated from the crystal structure by counting the number of heavy-atom contacts between residues $u$ and $v$, ${n_{uv}^{\text{nc}}}$, and determining whether a main-chain hydrogen bond exists; the hydrogen bond contribution is ${\alpha_{\text{hb}}}$ times that of a single heavy-atom contact.
Because native-centric models are known to over-stabilize helices~\cite{shimada2002ensemble,hubner2006understanding}, we weaken all energies associated with helical contacts by a factor ${\alpha_{\text{helix}}}$.
The bond energy formula is thus ${\epsilon_{uv} = -\left(\alpha_{\text{helix}}\right)^{\bm{1}_{uv}^{\text{helix}}} \left[n_{uv}^{\text{nc}} + \alpha_{\text{hb}}\bm{1}_{uv}^{\text{hb}}\right]}$, where~$\bm{1}_{uv}^{\text{hb}}$ indicates the presence of a hydrogen bond and $\bm{1}_{uv}^{\text{helix}}$ indicates a helical contact.
The constants ${\alpha_{\text{helix}}=5/8}$ and ${\alpha_{\text{hb}}=16}$ were chosen empirically to maximize the agreement with experiments on protein G (see SI \figref{fig:1igd} and \tabref{tab:phi1}).
The inverse temperature is then tuned to achieve a fixed free-energy difference between the unfolded and folded ensembles (see SI text).

\subsection*{Identification of cooperative substructures and transient states}
\label{sec:substructures}

We now seek to identify kinetically separated folding intermediates by examining the factors that give rise to free-energy barriers between microstates.
In the contact-graph model, all free-energy barriers are purely entropic, since the native contacts are assumed to be energetically favorable.
The most significant free-energy barriers arise from the formation of loops, which entail an entropic penalty of at least ${(b + 1)\mu/T}$ that is not immediately compensated by energetically favorable native contacts.
Once an initial loop has been formed, the recruitment of residues that are adjacent in the protein sequence may result in a net decrease in the free energy.
As a result, the model naturally gives rise to cooperative \textit{substructures}, i.e., sets of contacts that require the formation of a single loop and thus share a common free-energy barrier.
As in the helix--coil~\cite{zimm1959theory} and kinetic-zipper~\cite{dill1993cooperativity} models of peptide assembly, the sets of contacts comprising an individual substructure are typically bi-stable: either none of the contacts are made in the high-entropy state, or many contacts are required to compensate for the loss of conformational entropy in the low-energy state.

\begin{figure}
  \includegraphics{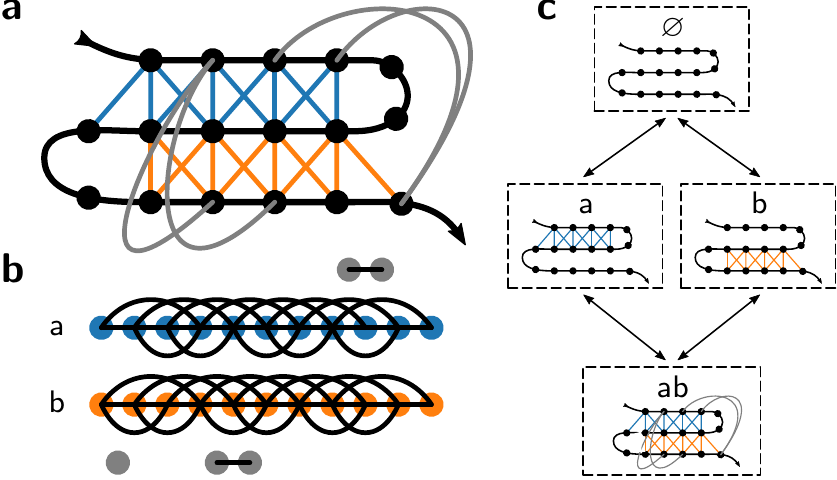}
  \caption{\textbf{Identification of substructures and topological configurations.}  (a)~An example contact graph, with contacts colored by substructure.  Each substructure requires the formation of one loop in the polymer backbone.  Unassigned contacts are shown in gray.  (b)~The `backbone-dual' graph, in which the vertices represent native contacts (see text).  (c)~A substructure is part of the current topological configuration if one or more of its contacts are formed in the largest structured region.  The unassigned contacts contribute to the stability of configuration \textsf{ab}.  Arrows indicate allowed transitions between topological configurations that differ by the addition or removal of one substructure.}
  \label{fig:substructures}
\end{figure}

We identify groups of contacts that constitute the distinct substructures of a contact graph using the following algorithm.
First, we find all pairs of contacts where the interacting residues are either identical or are adjacent on the polymer backbone; that is, two contacts ${(u,v)}$ and ${(r,s)}$ are linked if ${r-u \in \{-1,0,1\}}$ and ${s-v \in \{-1,0,1\}}$.
These pairs of contacts define a `backbone-dual' graph in which the vertices represent native interactions and the edges indicate adjacency along the polymer backbone (\figsref{fig:substructures}a and b).
We then decompose this graph into connected components, retaining only those components with at least six contacts in order to counter the minimum entropic cost of forming a Kuhn-length loop.
The role of contacts that are not assigned to substructures is discussed below.
While the substructures identified by this algorithm often align with elements of secondary structure, this does not have to be the case, since the substructures are defined purely on the basis of the three-dimensional native structure.

Advancing toward the folded state requires building up successive substructures, each of which is associated with a free-energy barrier.
A transition path must cross each of these barriers one-at-a-time, regardless of the precise order in which the contacts are formed.
These intermediate states can be described by a discrete set of \textit{topological configurations} that indicate the assembly of one or more substructures (\figref{fig:substructures}c).
In the remainder of this work, we simplify our analysis by tracking only the largest native-like cluster of residues.
As a result, each topological configuration refers to the assembly of a specific set of substructures within a single structured region.
The validity of this assumption is discussed in \secref{sec:simulations}.

Native contacts that are not assigned to substructures contribute to the stability of topological configurations that consist of multiple substructures in a single structured region.
For example, in \figref{fig:substructures}, the unassigned contacts shown in gray contribute to topological configuration \textsf{ab} but not to configuration \textsf{a} or \textsf{b}.
In cases where some residues do not participate in any of the identified substructures, we define a separate native configuration that contains all substructures plus the additional contacts involving these residues.
Because such residues do not contribute to any of the intermediate topological configurations, they do not affect the folding transition paths predicted by our theory; the contacts formed by these residues serve only to stabilize the native state.

Because of the significant free-energy barriers associated with loop formation, cooperative substructures are predicted to have long lifetimes compared to individual native contacts.
Furthermore, the free-energy barriers between topological configurations are expected to give rise to metastability: microstates that share the same set of loops can inter-convert rapidly, while transitions between topological configurations that differ by the addition or removal of one substructure occur on a much slower timescale.
These topological configurations therefore serve as an appropriate set of coarse-grained, transient states for analyzing the dynamics of protein-folding transition paths.

\subsection*{Prediction of folding transition paths}

Having established a structural definition of a transient state, we can now construct a rate matrix to describe stochastic transitions between the coarse-grained configurations.
First, we calculate the free-energy of each configuration, $F_i$, by summing over all microstates that conform to the topological configuration~$i$: ${F_i \equiv -\kB T \ln \sum_{\{g\}_i} \!\exp(-F_g / \kB T)}$.
The compatible microstates ${\{g\}_i}$ are those which have a single structured region and contain one or more contacts from each substructure comprising configuration~$i$.
This sum can be calculated efficiently via Monte Carlo integration using the technique described in \refcite{jacobs2015theoretical} (see SI text).
This calculation also yields the equilibrium probability of contact formation within each topological configuration, ${\langle \bm{1}_{uv} \rangle_i \equiv \sum_{\{g\}_i}\! \bm{1}_{uv}(g) \exp(-F_g / \kB T)}$.
As we shall demonstrate, the most probable microstates within a topological configuration may not form all possible contacts.

We then calculate the free-energy barriers, ${\Delta F_{i \rightarrow j}^\dagger}$, between topological configurations $i$ and $j$ that differ by the addition or removal of one substructure.
We consider two mechanisms of substructure addition: either the formation of a new loop via a single contact or the consolidation of a pre-formed substructure with the existing structured region.
The former mechanism is applicable when the added substructure shares residues with substructures in the existing structured region.
In contrast, the latter mechanism is applicable when the added substructure and the existing structured region have no residues in common but nevertheless form contacts in the native structure.
In both cases, we calculate the mean-field probability of forming an initial contact with one or more residues of the new substructure, assuming that the existing structured region is in local equilibrium.
The details of these calculations, which take into account fluctuations within each topological configuration, are provided in the SI text.

Finally, we construct a rate matrix to describe transitions between topological configurations.
The dimensionless rates $k_{ij}$ obey detailed balance and are assumed to follow from the Metropolis criterion,
\begin{equation}
  k_{ij} =
  \begin{cases}
    \exp\!\left[\min\!\left(0, -\frac{\Delta F_{i \rightarrow j}^\dagger}{\kB T} \right)\!\right] \!\!& \text{if } i,j \text{ adjacent}, \\
    -\sum_{j' \ne i} k_{ij'} & \text{if } i = j, \\
    0 & \text{if } i,j \text{ not adjacent.}\!
  \end{cases}
  \label{eq:k}
\end{equation}
From this rate matrix, it is straightforward to obtain ensemble-averaged properties of transition paths between the unfolded and folded ensembles using transition-path theory~\cite{metzner2009transition}.
Of particular interest are the the commitment probabilities, $\pfold(i)$~\cite{du1998transition}, and the folding fluxes, $f_{ij}$, between adjacent configurations.
In addition, we can predict folding intermediates by calculating the average time spent in each configuration within the transition-path ensemble.
Details are provided in the SI text.

\section{Results}
\label{sec:results}

\subsection*{Proteins fold via a sequence of transient states}

\begin{figure*}
  \includegraphics{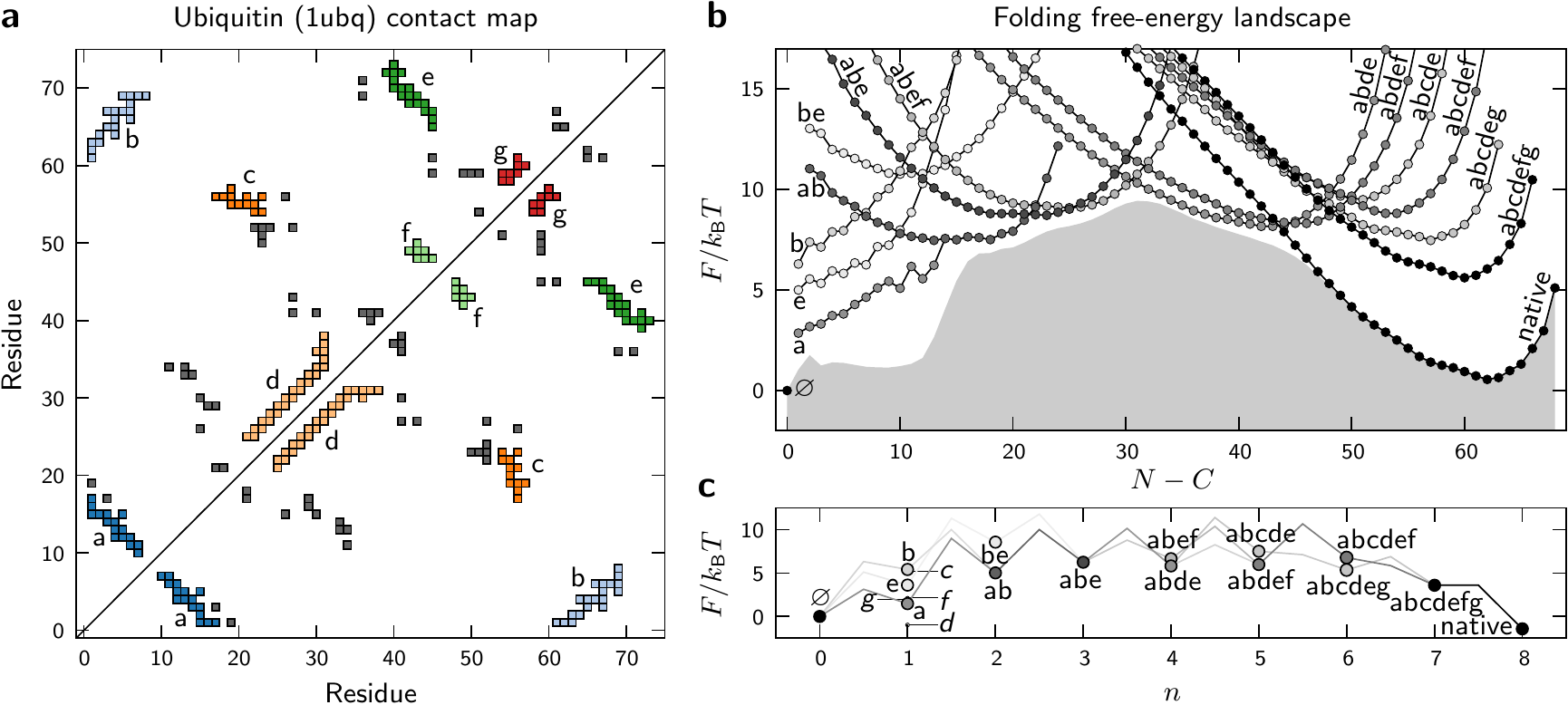}
  \vskip-1ex
  \caption{\textbf{Predicted folding free-energy landscapes for ubiquitin.} (a) The contact map obtained from the crystal structure of ubiquitin (protein data-bank entry 1ubq) indicating the discrete substructures \textsf{a}--\textsf{g} described in \secref{sec:substructures}.  (b) The free energy of each topological configuration as a function of the total number of interacting residues, $N$; the number of structured regions, $C$, is one for all configurations except the unfolded state, $\varnothing$, where ${C = 0}$.  The shaded region shows the one-dimensional free-energy profile.  (c) The free energy of each topological configuration as a function of the number of assembled substructures,~$n$.  All free energies are calculated relative to the state $\varnothing$, and the inverse temperature is tuned to achieve equal stabilities of the native and unfolded ensembles.  The shading indicates the fraction of the net folding flux through each configuration.  Only configurations with at least 10\% of the net folding flux are shown, except in panel \textbf{c}, ${n = 1}$, where all substructures are labeled.}
  \label{fig:landscapes}
\end{figure*}

Free-energy calculations support the interpretation of the substructures identified in \secref{sec:substructures} as the minimal cooperative units on a folding transition path.
As an example, we present calculations for ubiquitin in \figref{fig:landscapes}; its seven substructures are indicated on the contact map in \figref{fig:landscapes}a.
When plotted as a function of the total number of interacting residues, $N$, we find that every topological configuration is associated with a single local free-energy minimum (\figref{fig:landscapes}b).
Single-substructure configurations are typically unstable, as the free energy increases with the number of interacting residues.
In contrast, the energetically favorable native contacts in multiple-substructure configurations more than compensate for the loss of conformational entropy due to loop formation.
However, at the local minimum in each of these configurations, the polymer is unlikely to form all possible native contacts for entropic reasons: there are many more partially assembled microstates, and some residues make too few native contacts to offset the entropic cost of ordering completely.
As a result, the free-energy minimum typically occurs at a value of $N$ that is less than the maximum number of residues in each configuration.
Because of this competition between stabilizing native contacts and various entropic contributions, the locations of these free-energy minima are \mbox{temperature-dependent}.

Plotting the free-energy landscape as a function of the number of assembled substructures more clearly shows the free-energy barriers between adjacent topological configurations (\figref{fig:landscapes}c).
Microstates belonging to different topological configurations are kinetically separated by at least one entropic barrier and cannot inter-convert rapidly.
The existence of significant free-energy barriers between unimodal free-energy basins supports the assertion that the topological configurations constitute transient states on the transition paths between the completely unfolded and native states.
Alternate pathways may be traversed, depending on the order in which the free-energy barriers between configurations are crossed.
Yet in general, we find that only a small number of parallel pathways contain the vast majority of the reactive flux between the unfolded and native states.
In \figsref{fig:landscapes}b and c, the shading of each topological configuration indicates the fraction of the net folding flux, ${f_{ij}^+ \equiv \max(f_{ij}-f_{ji},0)}$, passing through that configuration on folding transition paths; the many other configurations with negligible net folding flux are not shown.

\subsection*{Free-energy landscapes predict a common folding mechanism}

These multi-modal free-energy landscapes point to a common folding mechanism.
As expected on the basis of \eqref{eq:Fg}, our free-energy calculations indicate that there are no significant barriers separating microstates within individual topological configurations.
Instead, the relevant barriers are found \textit{between} topological configurations.
These landscapes thus predict that folding proceeds by the step-wise consolidation of cooperative structures within a single structured region.
The transition state on a folding pathway is reached upon the formation of a specific set of substructures, after which all subsequent barriers on the pathway are lower in free energy and folding can proceed downhill to the native state.

\begin{figure*}
  \includegraphics{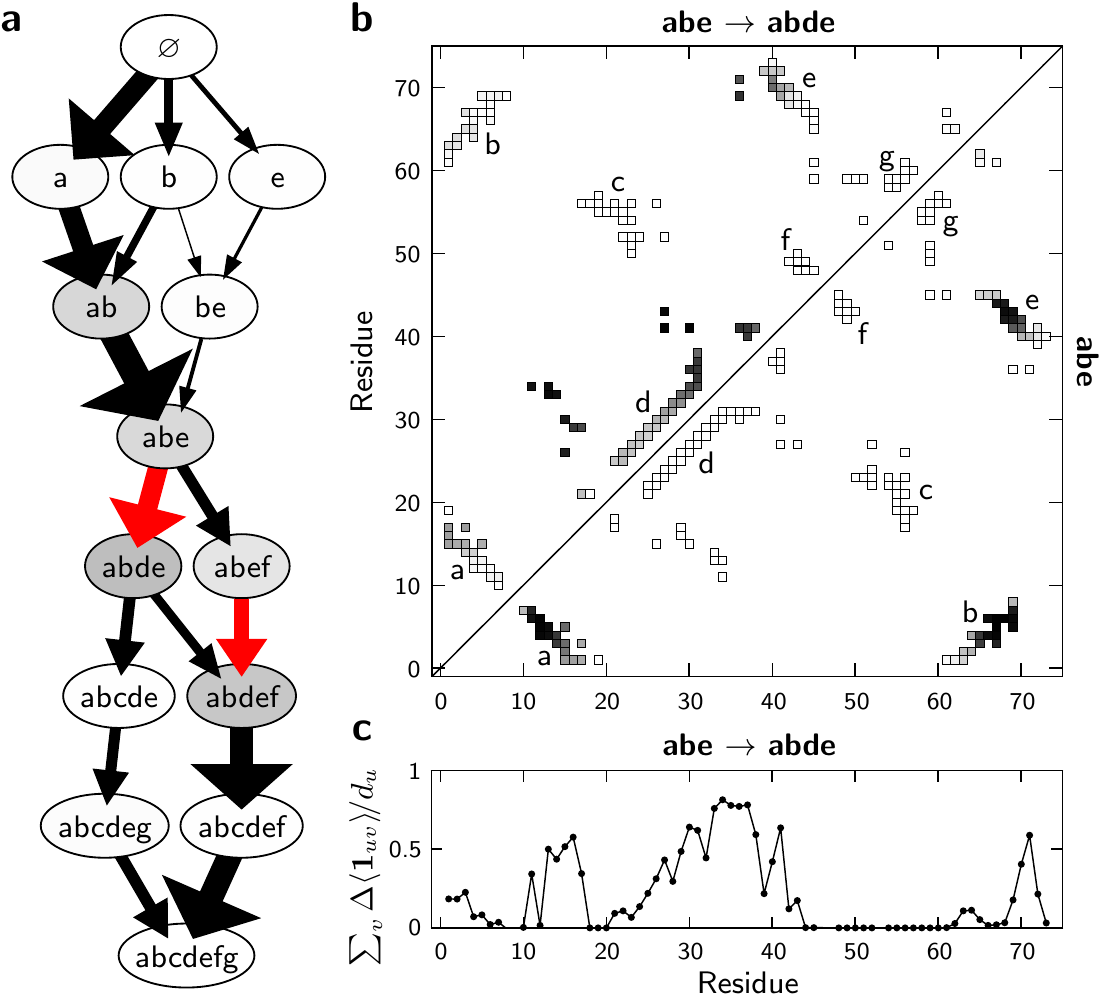}
  \caption{\textbf{Specific contacts are formed at the rate-limiting step on the folding transition paths of ubiquitin.} (a) The folding network of ubiquitin, showing the topological configurations containing at least 10\% of the net folding flux (see text).  (b) Below the diagonal, the equilibrium distribution of native contacts in topological configuration \textsf{abe}.  Above the diagonal, the difference between the equilibrium contact distributions of configurations \textsf{abe} and \textsf{abde}.  Black indicates a probability of one, while white indicates zero.  (c) The difference in equilibrium contact formation, ${\Delta\langle\bm{1}_{uv}\rangle}$, between configurations \textsf{abe} and \textsf{abde}, averaged over each residue.  The total number of native contacts made by residue $u$ is $d_u$.  A small number of essential long-distance contacts, primarily involving residues 13--17, 27--41 and 69--71, are formed at the transition between these configurations.}
  \label{fig:network}
\end{figure*}

In order to preserve the kinetic connectivity of the transient states, the folding free-energy landscape is best represented by a network of the discrete topological configurations.
In \figref{fig:network}a, we show all configurations containing at least 10\% of the net folding flux.
Arrows indicate the net folding flux between configurations, while the shading indicates the fraction of the total transition-path time spent in each configuration.
The transitions that pass through the rate-limiting step, from which the protein has an equal probability of folding or unfolding, are highlighted.
This kinetic network shows that substructures tend to assemble in a remarkably well-ordered sequence, despite the stochastic nature of the transitions between transient states.
It is important to note that this ordering is not dictated simply by the stability of the isolated structures: the sequence of events on folding transition paths does not match the ranking of the substructure free energies (${n = 1}$) in \figref{fig:landscapes}c.
Instead, the most likely pathway depends on the stability of the intermediate configurations and the barriers between them, which in turn depend on the contacts between \mbox{substructures.}

Although many proteins are commonly described by two-state kinetics, our analysis indicates that folding transition paths may have greater kinetic complexity due to the presence of transient, high-free-energy folding intermediates.
For comparison, a one-dimensional profile showing the free energy as a function of the number of interacting residues is shown in \figref{fig:landscapes}b.
In contrast to our approach, this representation of the folding landscape does not distinguish among microstates in directions orthogonal to the order parameter and consequently hides the barriers that prevent microstates with similar numbers of interacting residues from inter-converting rapidly.
Decomposing the landscape into topological configurations provides more detailed insights into the folding free-energy barrier and the trade-off between native-contact formation and the loss of conformational entropy.
In particular, our analysis shows that a specific set of loops in the polymer backbone must be formed in order for subsequent native contacts to lower the free energy as folding progresses toward the native state.

\subsection*{Specific contacts are formed at the rate-limiting transition}

\figref{fig:network}a shows that the assembly of topological configuration \textsf{abde} or \textsf{abdef} is required for ubiquitin to reach the folded ensemble.
Common to both of the highlighted transitions is the consolidation of the helix (substructure~\textsf{d}) with a partially formed $\beta$-sheet (substructures~\textsf{a}, \textsf{b} and \textsf{e}); the final hairpin of the $\beta$-sheet (substructure~\textsf{f}) is optional and thus largely irrelevant.
This analysis provides a clear mechanistic description of the essential rate-limiting event on a folding transition path.
In addition, our analysis predicts that the majority of the transition-path time is spent in the metastable configurations just before and after the transition, configurations~\textsf{ab}--\textsf{abdef}.

Importantly, this approach allows us to distinguish between the native contacts that are prerequisite for reaching the transition state and those that are formed precisely at the rate-limiting step.
As illustrated in \figref{fig:network}b, a relatively small number of native contacts are involved in the rate-limiting step on ubiquitin's folding pathway.
Shown below the diagonal in this plot is the contact distribution in the pre-transition configuration \textsf{abe}, assuming local equilibrium in this metastable state.
Not all contacts within the three contributing substructures are equally probable; in particular, residues near the extremities of the $\beta$-sheet are more likely to be disordered.
In order to determine the contacts that are formed upon the incorporation of the helix into the largest structured region, we subtract the union of the contact distributions of configuration \textsf{abe} and the isolated substructure \textsf{d} from the post-transition configuration \textsf{abde}.
We find that a specific set of approximately 15 long-range contacts between the helix and partial $\beta$-sheet are essential for the rate-limiting transition.
The residue-averaged contact differences (\figref{fig:network}c) indicate that these specific contacts primarily involve residues 13--17, 27--41 and 69--71.
As we shall show below, this distribution of rate-limiting contacts is significantly different from the complete set of contacts present at the transition state.

\subsection*{Comparison with atomistic Molecular Dynamics simulations}
\label{sec:simulations}

\begin{figure*}
  \includegraphics{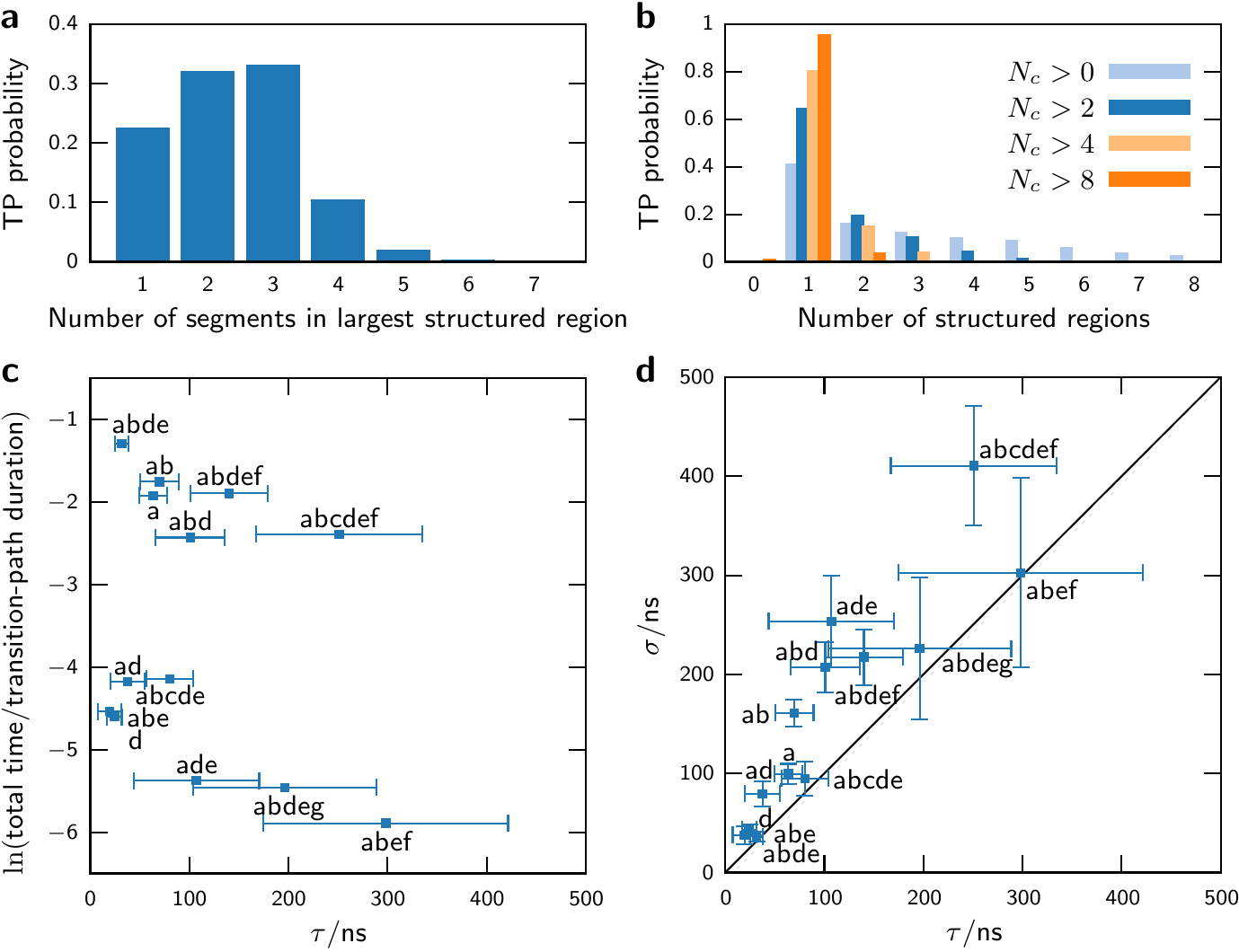}
  \vskip-1ex
  \caption{\textbf{Verification of the assumptions and predictions of the theory using atomistic simulations (see~\refcite{piana2013atomic}).}  (a)~For $\sim$50\% of the transition-path (TP) duration, more than two native-like segments are formed in the largest structured region.  (b)~Histograms of the number of distinct structured regions with a minimum number of residues in the transition-path ensemble.  (c)~The fraction of the total transition-path time spent in each topological configuration.  (d)~The mean, $\tau$, versus standard deviation, $\sigma$, of the topological-configuration lifetimes.  The line ${\sigma=\tau}$ is indicative of an exponential waiting-time distribution.}
  \label{fig:assumptions}
\end{figure*}

The accuracy of these predictions can be tested by comparison with atomistic Molecular Dynamics simulations.
For this purpose, we obtained unbiased simulation trajectories of the reversible folding and unfolding of wild-type ubiquitin from Shaw and co-workers~\cite{piana2013atomic}.
We shall focus our attention on the native contacts formed during the $\sim 1$--$10~\mu$s-long transition paths (two folding and eight unfolding) that were captured from six independent simulations.
The details of our analysis are provided in the SI text; the Molecular Dynamics simulations are described in \refscite{piana2013atomic,lindorff2011fast}.

We first tested the underlying assumptions of our theoretical approach.
\figref{fig:assumptions}a shows a histogram of the number of segments within the largest structured region in the ensemble of transition-path structures.
The segments here are defined as stretches of sequential residues forming native contacts, with the additional constraint that each segment is separated by at least $b$ non-interacting residues.
This histogram clearly shows that a single or double-sequence approximation, i.e., assuming one or two native-like segments, is inadequate.
In contrast, we verified that modeling only the largest structured region is sufficient to describe most of the transition-path ensemble.
In \figref{fig:assumptions}b, we plot the probability of finding one or more structured regions, each containing a minimum number of residues $N_c$, on a transition path.
If we ignore all native-like clusters containing eight or fewer residues, then we find that the assumption of a single structured region is valid for over 95\% of the (un)folding trajectories.

Next, we calculated the lifetimes of the predicted transient states on the observed transition paths.
As in our theoretical approach, we identified the topological configuration in the simulation trajectories by determining which substructures are at least partially formed within the largest structured region.
We then calculated the mean, $\tau$, and standard deviation, $\sigma$, of the distribution of lifetimes for all visits to each topological configuration.
In \figref{fig:assumptions}c, we plot the fraction of the total transition-path time spent in each configuration versus its mean lifetime.
We find that the three most populated configurations (\textsf{abde}, \textsf{ab}, and \textsf{abdef}) agree with the predictions shown in \figref{fig:network}a.
Meanwhile, the average lifetimes of all visited transient states range from 20 to 300 ns, considerably longer than the timescale for native-contact formation.
Finally, \figref{fig:assumptions}d shows that the coefficient of variation of the lifetimes, ${\sigma/\tau}$, is close to unity for most configurations.
This is indicative of an exponential distribution of waiting times, which supports our prediction that the barrier-separated configurations constitute metastable states.

\begin{figure}
  \includegraphics{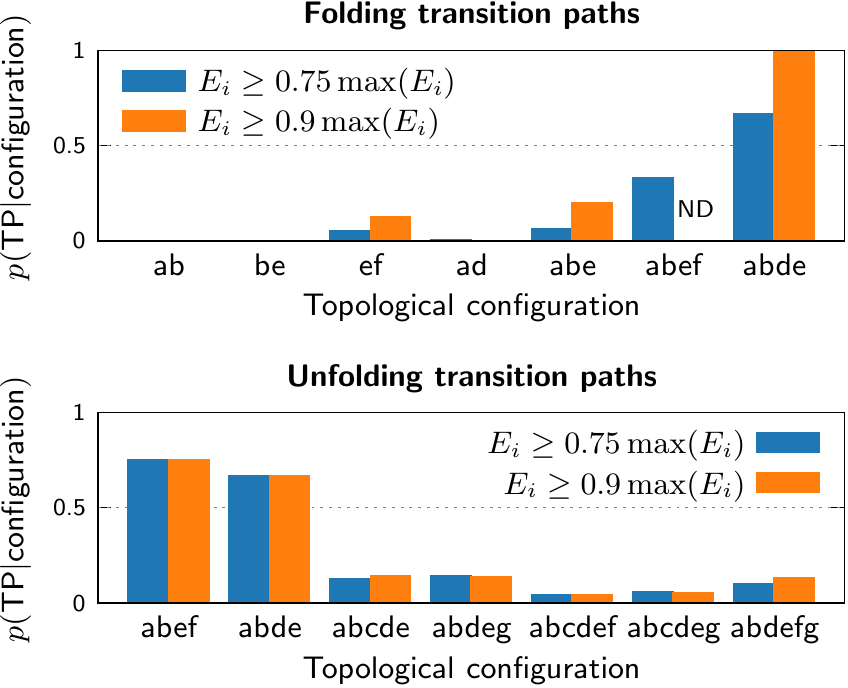}
  \caption{\textbf{Commitment probabilities for transient states in atomistic simulations.}  The probability of being on a transition path given that an excursion from the unfolded (top) or native (bottom) ensemble either reaches or disrupts the indicated topological configuration, respectively.  Only excursions that achieve a minimum fraction of the total number of contacts in a topological configuration, ${\max(E_i)}$, are counted.  No data is available in the case of configuration \textsf{abef} for the stricter condition ${E_i \ge 0.9\max(E_i)}$, since no qualifying events were observed in the available simulation trajectories.}
  \label{fig:commit}
\end{figure}

Having verified our fundamental assumptions and the most general predictions of our theory, we then assessed the accuracy of our predictions regarding the rate-limiting step of the folding reaction.
We identified all excursions away from the free-energy minima of the unfolded and folded ensembles in the simulation trajectories and counted the number of excursions that reached each topological configuration starting from either the unfolded, U, or folded, F, ensemble.
Transitions were only counted if a minimum fraction of the total number of contacts, ${\max(E_i)}$, are formed in configuration~$i$.
We then calculated the commitment probability for each configuration, i.e., the probability of being on a transition path given that a specific topological configuration is reached, using the Bayesian formula
\begin{equation}
  p{(\text{TP}\,|\,\text{U\!/F} \rightarrow i) = \frac{n_{\text{TP}} \times p(\text{U\!/F}\rightarrow i\,|\,\text{TP})}{n_{\text{U\!/F}\rightarrow i}}},
\end{equation}
where $n_{\text{TP}}$ is the number of folding or unfolding transition paths, $p(\text{U/F}\rightarrow i\,|\,\text{TP})$ is the probability of reaching configuration~$i$ on a folding or unfolding transition path, and $n_{\text{U/F}\rightarrow i}$ is the total number of excursions that reached configuration~$i$.
The results of this analysis are presented in \figref{fig:commit}.

As predicted, the probability of folding surpasses 50\% once configuration \textsf{abde} is reached from the unfolded ensemble; with the stricter criterion ${E_i \ge 0.9\max(E_i)}$, this probability increases to 100\%.
The necessary precursors to this transition, including the assembly of substructures \textsf{a}, \textsf{b} and \textsf{e}, have considerably smaller commitment probabilities.
We also find that disrupting configuration \textsf{abde} increases the probability of unfolding above 50\% for excursions starting from the folded ensemble.
Despite the limited statistics from the available simulation trajectories, these results lend strong support to our predictive theory.
This agreement is crucial because it demonstrates that our description in terms of transient states can provide mechanistic insights into the rate-limiting events on the transition paths of topologically complex proteins.

\subsection*{Comparison with kinetic measurements}

Experimentally, the folding transition-state ensemble can be probed indirectly by perturbing interactions between residues.
The most commonly used techniques are $\phi$-value analysis~\cite{fersht1992folding}, which compares changes in the rate of folding to changes in the equilibrium constant due to single-residue point mutations, and $\psi$-value analysis~\cite{krantz2004discerning}, which applies an analogous strategy to pairwise contacts between solvent-exposed residues.
While $\phi$ and $\psi$-values do not test our theory directly --- for instance, they cannot distinguish the rate-limiting contacts from prerequisite contacts at the transition state, nor can they provide detailed information on transition-path dynamics --- they remain the only experimental techniques for which consistent data exist for a large number of proteins.

In order to compare our model with experimental measurements, we calculate $\phi$ and $\psi$-values due to energetic perturbations in the small-perturbation limit,
\begin{eqnarray}
  \psi_{uv} &=& \left. \Delta_{uv} \! \left( \ln \kfold^{-1} \right) / \left( \Delta_{uv} F_{\text{native}} / \kB T \right) \right|_{\epsilon'_{uv}\! - \epsilon_{uv} \rightarrow 0},\;
  \label{eq:psi} \\
  \phi_u &=& \sum_v \psi_{uv} / d_u,
  \label{eq:phi}
\end{eqnarray}
where $\kfold$ is the folding rate calculated from transition-path theory, ${\Delta_{uv}}$ indicates the change due to a perturbation in the contact energy ${\epsilon_{uv} \rightarrow \epsilon'_{uv}}$, and $d_u$ is the number of contacts made by residue $u$ in the native state.
In $\phi$-value comparisons, we consider only mutations to alanine or glycine; in cases where data for both mutations are available, we choose the substitution that is chemically most similar to the wild-type residue at that position.
We also leave $\phi$-values that are negative or significantly greater than unity out of the comparison (see SI text).

\begin{figure*}
  \includegraphics{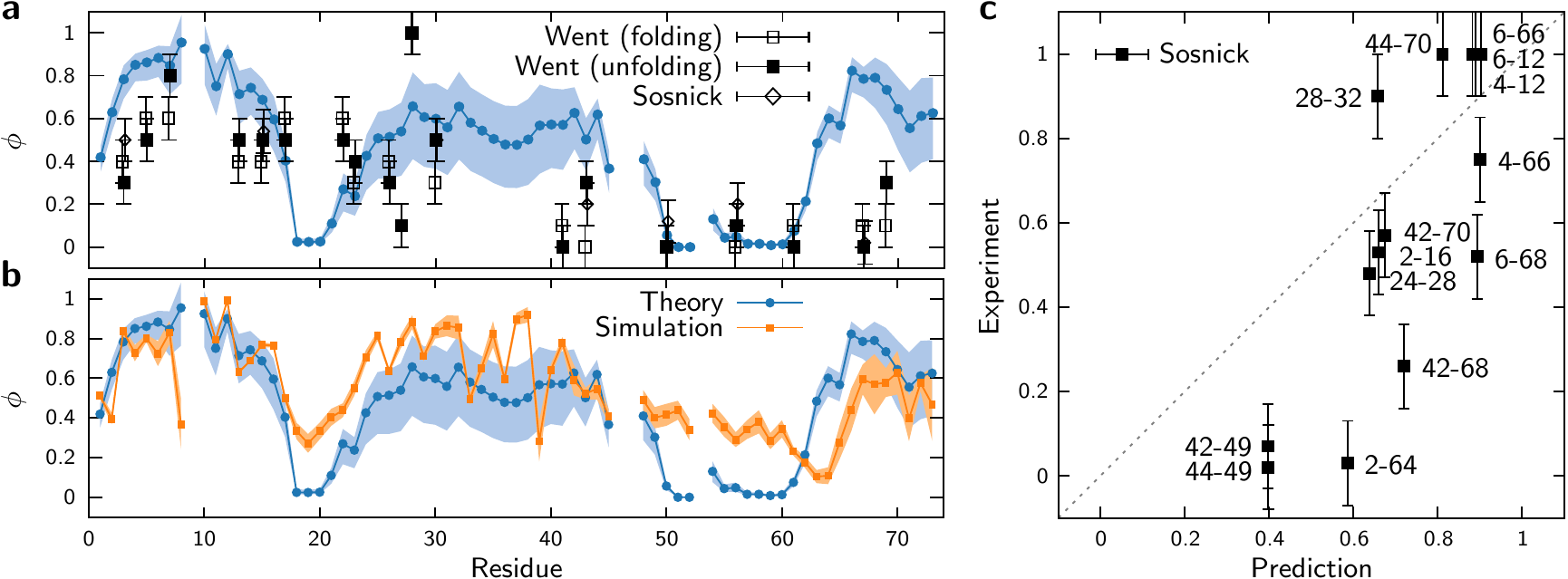}
  \caption{\textbf{Comparison of $\phi$ and $\psi$-values for ubiquitin.}  (a) Comparison of predicted $\phi$-values in the small-perturbation limit with three sets of experimental measurements: Went et al.~(folding and unfolding), \refcite{went2005ubiquitin}, and Sosnick et al., \refcite{sosnick2004differences}.  Circles indicate predictions assuming ${\Delta F_{\text{native}} = 0}$; the light-blue region indicates the range of predictions from ${\Delta F_{\text{native}} = 2~\kB T}$ (upper limit) to ${-2~\kB T}$ (lower limit).  Predictions are not shown for residues that do not form native contacts.  (b) Comparison with $\phi$-values calculated from atomistic simulations.  The light-orange range reports an estimate of the variability across individual transition paths (see SI text).  (c)  Comparison of predicted and experimental $\psi$-values from Sosnick et al.,~\refcite{sosnick2004differences}.}
  \label{fig:phi_psi_comparison}
\end{figure*}

In \figref{fig:phi_psi_comparison}, we show the agreement between the predicted $\phi$ and $\psi$-values and three experimental measurements for ubiquitin.
Calculating the $\phi$-value predictions under conditions of equal folded and unfolded populations (see SI text), we obtain a correlation coefficient ${R = 0.43}$ and p-value ${p = 0.063}$ with the unfolding data of \refcite{went2005ubiquitin}.
To get an idea of the variability in our predictions due to changes in the native-state stability, we also plot the predicted range of $\phi$-values due to stabilizing or destabilizing the native state by ${2~\kB T}$.
This agreement is reasonable considering that many mutations perturb the energy of the transition state by several $\kB T$.
The correlation between the predicted and experimental $\psi$- values is considerably stronger, with ${R = 0.80}$ and ${p = 0.00061}$.
There is less ambiguity in the latter comparison, because the experimental perturbations are intended to affect only a single native contact and are reported in the small-perturbation limit.
We also compare our predictions with $\phi$-values calculated from the atomistic simulations following the transition-path ensemble method of \refcite{best2016microscopic} and the native-contact definition used in \figsref{fig:assumptions} and \ref{fig:commit}.
Here we find that the theory--simulation and simulation--experiment correlations for $\phi$-values are similar (${R = 0.60,}~{p = 4.5\times10^{-8}}$ and ${R = 0.51},~{p = 0.024}$, respectively); however, the agreement between simulation and experiment is weaker for $\psi$-values (${R = 0.48},~{p = 0.080}$).
Notably, both the theoretical predictions and the simulation results indicate a more pronounced role for the C-terminus in the transition-state ensemble than is apparent from the experimental $\phi$-values (\figsref{fig:phi_psi_comparison}a and b).

To examine the generality of our predictions, we have also calculated $\phi$ and $\psi$-values for comparison with experiments on an additional 14 proteins.
Overall, we find good agreement, indicating that our native-centric model captures the essential physics of folding across a wide variety of proteins with 50 or more amino acids (\tabref{tab:phi_psi}).
Detailed case studies for protein~G (1igd), protein~L (1k53), chymotrypsin inhibitor~2 (2ci2), cold-shock protein (1csp) and an SH3 domain (1shg) are provided in the SI text; complete details of all mutations tested are provided there as well.
We find that the agreement between our predictions and experiments is generally better for $\psi$-values than $\phi$-values and is worst for small helix bundles, such as the engrailed homeodomain (1enh), which are known to have heterogeneous folding pathways that are highly sensitive to the force field used in computer simulations~\cite{best2016microscopic,piana2011robust}.
In fact, the greatest source of uncertainty in making these comparisons is the sensitivity of the predicted $\phi$ and $\psi$-values to the native-contact energies, and, consequently, the relative stabilities of the substructures.
We caution that the calculated correlation coefficients and p-values are affected by correlations in the $\phi$ and $\psi$-values of neighboring residues and the choice of mutations for experimental characterization.
Nevertheless, these results indicate that the predictions of our theory are compatible with the available experimental data on a diverse set of proteins.

\section{Discussion}

We have introduced a theory to predict the detailed kinetics and intermediate states on protein-folding transition paths.
We have shown that the folding of topologically complex proteins follows a predictable sequence of transitions between transient states, which can be identified directly from the native structure.
While our approach has been developed using a discrete, native-centric model of globular proteins, our conclusions are broadly applicable to the self-assembly of polymers with specific interactions, such as non-coding RNA~\cite{eddy2001non} and DNA origami~\cite{rothemund2006folding}.

\subsection*{Physical explanation for the emergence of `foldons'}

Our analysis shows that there is a natural level of resolution for describing transition-path dynamics.
Although all macromolecular transition paths are heterogeneous when examined in sufficient detail, modeling the assembly and disassembly of discrete substructures fully captures the long-timescale motions and metastable states on folding pathways.
In addition, this ability to predict transient states on the basis of a protein's native structure alleviates the need for contiguous-sequence approximations that are not justified for proteins with complex native topologies.

Many lines of evidence, including hydrogen-exchange~\cite{maity2005protein}, metal-binding kinetics at bi-histidine sites~\cite{bosco2009metal} and single-molecule pulling experiments~\cite{mickler2007revealing}, support the existence of transient, high-free-energy folding intermediates composed of cooperative units that are often referred to as `foldons'~\cite{lindberg2007malleability,englander2014nature}.
In fact, sequential folding through a series of intermediates was proposed in some of the earliest models of protein-folding~\cite{ptitsyn1973stagewise,karplus1976protein}.
Our theory predicts that these transient states emerge directly from the topology of the native state.
We have further shown that the cooperativity among these groups of native contacts is a consequence of the central role of loop formation in protein folding, which gives rise to entropic barriers between transient states.
While these cooperative units are most easily identified in the context of a native-centric model, the appearance of structurally defined metastable states in atomistic simulations supports the generality of this finding.

\begin{table}[t]
  \centering
  \begin{minipage}[t]{4.25cm}
  \begin{tabular}{clll}
    \textbf{$\phi$-values:} & $n$ & $R$ & $p$ \\
    \hline
    1enh & 11 & 0.14 & 0.69 \\
    1igd & 20 & 0.80 & 0.000023 \\
    1shg & 10 & 0.70 & 0.024 \\
    1k53 & 37 & 0.36 & 0.031 \\
    2ci2 & 32 & 0.42 & 0.018 \\
    1csp & 16 & 0.71 & 0.0041 \\
    1ubq & 19 & 0.43 & 0.063 \\
    1imp & 14 & 0.73 & 0.003 \\
    1tiu & 22 & 0.48 & 0.024 \\
    1btb & 21 & 0.44 & 0.045
  \end{tabular}
  \end{minipage}
  \begin{minipage}[t]{4.25cm}
  \begin{tabular}{clll}
    \textbf{$\phi$-values:} & $n$ & $R$ & $p$ \\
    \hline
    1fkb & 21 & 0.65 & 0.0015 \\
    1rnb & 12 & 0.60 & 0.038 \\
    3chy &  7 & 0.91 & 0.0044 \\
    2vil & 17 & 0.43 & 0.088 \\
    \rule{0pt}{6ex} 
    \textbf{$\psi$-values:} &  $n$ &    $R$ &    $p$ \\
    \hline
    1igd &  8 & 0.69 & 0.059 \\
    1k53 &  7 & 0.93 & 0.022 \\
    1ubq & 14 & 0.80 & 0.00061 \\
    2acy &  8 & 0.71 & 0.048
  \end{tabular}
  \end{minipage}
  \caption{\textbf{Comparison of predicted and experimental $\phi$ and $\psi$-values for a diverse set of proteins.}  For each protein, identified by its protein data-bank entry, we list the number of data points, $n$, the Pearson correlation coefficient, $R$, and the associated p-value, $p$.  Note that the 1igd $\phi$-values were used in the parameterization of the empirical two-parameter potential (see \secref{sec:theory}).  Complete details and accompanying figures are provided in the SI text (see SI \tabsref{tab:phi1}--\ref{tab:psi} and \figsref{fig:phi1}--\ref{fig:psi}).}
  \label{tab:phi_psi}
\end{table}

\subsection*{Ordered pathways are determined by the native-state structure}

Although protein-folding is a stochastic process, the most probable transition paths tend to follow a small number of distinct pathways.
Calculations for a structurally diverse set of examples (see SI \figsref{fig:1igd}--\ref{fig:1csp}) show that the dominant folding pathways are highly predictable when analyzed at the level of discrete substructures.
However, the order in which the substructures assemble is not determined by their stabilities in isolation.
Instead, the lowest-free-energy path through the folding landscape depends on both the stabilities of composite assemblies of multiple substructures and the barriers between these intermediate states.

This description in terms of transient states provides a detailed explanation for the origin of the folding free-energy barrier.
In the unfolded ensemble, the individual substructures tend to be unstable because the native contacts do not completely compensate for the loss of configurational entropy.
The lowest-free-energy folding pathway requires the assembly of a specific set of native-like loops in the polymer backbone, which then allows for the formation of stabilizing native contacts.
In particular, long-distance contacts~\cite{vendruscolo2002small,dokholyan2002topological} that connect the discrete substructures are most likely to form during a transition between topological configurations.
Because the ensemble of transition paths passes through a network of intermediates~\cite{rao2004protein,gfeller2007complex}, a folding reaction may be poorly described by a single order parameter.
In contrast to one-dimensional free-energy projections, coarse-graining on the basis of the topology of the polymer backbone preserves the kinetic connectivity of the complete folding landscape.

\subsection*{Mechanistic description of a folding reaction}

Our theory provides a mechanistic description of protein-folding transition paths by identifying the crucial event that must occur in order for a protein to fold to its native state.
The ability to predict the contacts that are formed at each step along the folding pathway is a key insight that is difficult to discern from kinetic measurements alone.
Whereas $\phi$ and $\psi$-values can, in principle, report the set of contacts that are formed at the transition state, our approach is able to distinguish which of these contacts are responsible for commitment to the folded ensemble.
In fact, many of the residues that form such crucial contacts, and are thus essential to the mechanism of folding, are found to have low to moderate $\phi$-values.
This is largely a consequence of averaging over all native contacts involving the residue of interest, only some of which may be formed at the transition state.
Core-facing residues that form a large number of stabilizing contacts in the native state are particularly likely to have low $\phi$-values for this reason~\cite{shakhnovich1997theoretical,hubner2004commitment}.
Other authors have noted that misleadingly low $\phi$-values from destabilizing mutations can result from structural relaxation in the transition state~\cite{sosnick2004differences} or redirection of the transition-path ensemble through parallel pathways~\cite{best2016microscopic}.

It is important to note that our predictions and the agreement with kinetic measurements are affected by the native-contact energies.
While the two-parameter empirical potential that we have used here is insufficient to capture all aspects of the interatomic interactions, we nevertheless achieve similar or greater accuracy in $\phi$ and $\psi$-value predictions to that of atomistic simulations (see, e.g.,~\refcite{best2016microscopic}).
This aspect of our theoretical predictions could be improved by increasing the complexity of the empirical potential and tuning the parameters by comparison with detailed simulation data.
Nevertheless, we expect that the general features of the predicted transition paths, including the metastability of structurally defined transient states, will remain unchanged.

\section{Conclusion}

In summary, we have developed an approach to predict protein-folding transition paths and high-free-energy intermediate states using a discrete native-centric model.
Our theory yields detailed, mechanistic insights into protein folding without the use of computationally expensive simulations.
Fundamentally, this advance relies on the physically realistic restrictions placed on the polymer configurations in our model, a crucial aspect that differs significantly from earlier efforts~\cite{munoz1999simple,alm1999prediction,galzitskaya1999theoretical}.

Beyond proteins, our theory can be applied more generally to polymers with specific interactions, such as non-coding RNA and DNA origami, where the ability to distinguish among kinetically separated pathways is essential for describing complex folding reactions.
The model that we have presented here is transferable to a variety of such systems due to the similar underlying physics of self-assembling structures that are built around polymer backbones and stabilized by native contacts.
We anticipate that this work will open up new avenues for addressing poorly understood aspects of protein-folding kinetics, including the molecular mechanisms of co-translational and chaperone-assisted folding.

\begin{acknowledgments}
  The authors would like to acknowledge Adrian Serohijos, Michael Manhart and David de Sancho for many insightful discussions and William Eaton for helpful comments on the manuscript.
  We are grateful to DE Shaw Research for providing access to the atomistic simulation trajectories.
  This work was supported by NIH grants R01GM068670 and F32GM116231.
\end{acknowledgments}


\renewcommand{\thefigure}{S\arabic{figure}}
\renewcommand{\thetable}{S\arabic{table}}
\renewcommand{\theequation}{S\arabic{equation}}
\renewcommand{\thesection}{S\arabic{section}}
\setcounter{figure}{0}
\setcounter{table}{0}
\setcounter{equation}{0}
\setcounter{section}{0}

\clearpage

\onecolumngrid
\section*{Supplementary Information for ``Structure-based prediction of protein-folding transition paths''}
\vskip0.5ex
\twocolumngrid

\section{Contact-graph model}

\subsection{Allowed microstates}
\label{sec:allowed_microstates}

In this section, we describe the allowed microstates of the contact-graph model using the language of graph theory.
The microstate that corresponds to the completely folded polymer, i.e., the configuration in which all possible contacts are formed, is denoted by the graph~$G$.
The vertices of this graph correspond to residues, while the edges indicate native contacts.
The vertices ${\{u\}}$ are labeled by their positions on the polymer backbone, i.e., ${u \in \{1,\ldots,L\}}$, where $L$ is the total number of residues in the chain.
We denote the set of all allowed subgraphs by ${\{g\}}$ and the set of edges in a microstate $g$ by ${\mathcal{E}(g)}$.
Because $g$ is a subgraph of $G$, every edge in ${\mathcal{E}(g)}$ is also an edge in ${\mathcal{E}(G)}$.
Only residues that form one or more contacts are represented by vertices in~$g$; this set of vertices is denoted by the set ${\mathcal{V}(g)}$.
The set of connected components of $g$ is ${\mathcal{C}(g)}$, and the edge and vertex sets of a connected component ${c \in \mathcal{C}(g)}$ are ${\mathcal{E}(g,c)}$ and ${\mathcal{V}(g,c)}$, respectively.

For each microstate, the associated graph of native contacts can be decomposed into a disjoint set of connected components (maximal subgraphs in which all pairs of vertices are connected by paths through the subgraph).
As described in the main text, the fact that the residues occupy non-overlapping finite volumes implies that many contacts must be correlated.
These correlations place restrictions on the combinations of contacts that can be simultaneously formed.
In the generation of a contact-graph model from a crystal structure, we have ignored contacts between residues that are separated by less than one Kuhn segment, $b$, in the polymer sequence; for consistency, we must therefore consider contacts involving sequences of residues that are shorter than one Kuhn segment to be correlated as well.
Consequently, we restrict the set of allowed microstates to those subgraphs that satisfy the following two rules:
\begin{enumerate}
\item Every connected component, ${c \in \mathcal{C}(g)}$, must be an induced subgraph of $G$.  This means that every edge $(u,v)$ in the connected component $c$ must appear in the subgraph $g$ if the vertices $u$ and $v$ are adjacent in the supergraph $G$.
\item Assume that two vertices ${v'\! > u'}$ belong to the same connected component $c$ and are separated by at most $b$ residues in the sequence, i.e., ${v'\! - u'\! \le b}$.  Then every intervening vertex $u$, i.e., ${u'\! < u < v'}$, must also be included in the connected component $c$ if an edge exists between $u$ and any vertex $v$ in $c$.
\end{enumerate}

\subsection{Loop entropy}

In \eqref{eq:Sloop}, we define a loop to be any contiguous sequence of non-interacting residues, with the exception of `bridge' segments (residues that, if removed, would break a polymer configuration given by a specific microstate into two non-interacting pieces).
For example, the microstate shown on the right in \figref{fig:graph}c contains two loops, 4--5--6--7 and 18, and one bridge segment, 11--12--13, where the residues are labeled starting from 1 at the top right of the figure.
In \eqref{eq:Sloop}, $r(l)$ is the end-to-end distance of loop $l$, expressed as a dimensionless multiple of the covalent backbone bond length; ${r = 0}$ if the residues at the loop ends form a native contact.

\subsection{Native-contact energies}

It is important to note that the native-contact energies are themselves free energies, since they depend on the average potential energy between two amino acids as well as solvent effects.
Here we assume that these attractive interactions are short-ranged and discrete, i.e., a contact is either completely formed or not present.
In a real polymer, there are likely to be other random interactions between residues.
Such nonspecific interactions contribute to the average energy of the ensemble of random coil configurations, which is taken to be the reference state for all free-energy calculations.
Consequently, the attractive interactions that are associated with specific contacts are, more precisely, associated with the \textit{differences} between the specific contact free energies and the average interaction energy between any pair of residues in the chain.
We assume that only these free-energy differences determine the folding pathways of the polymer.

The two-parameter empirical potential introduced in \secref{sec:theory} was manually tuned to achieve good agreement with the experimental $\phi$-values for protein G (1igd).
We verified that our values for the two adjustable parameters, ${\alpha_{\text{helix}}}$ and ${\alpha_{\text{hb}}}$, also result in close to optimal agreement with the experimental $\phi$-values for the $\alpha$/$\beta$ proteins 1k53, 1ubq and 2ci2.

\section{Monte Carlo free-energy calculations}

We compute free energies in this model using Monte Carlo integration.
This application of the Monte Carlo method is not a conventional simulation, as the sequence of microstates generated by our algorithm does not correspond to a physical folding trajectory.
Instead, the approach used here is simply an efficient means to integrate over the set of microstates with the same topological configuration.
(For a related application of this Monte Carlo technique, see \refcite{jacobs2015theoretical}.)
To do so, we first construct a Markov Chain to sample from the space of allowed subgraphs ${\{g\}}$.
We then use the Wang--Landau method~\cite{wang2001efficient} to calculate ${F_i}$, the free energy of all microstates in topological configuration~$i$.
Finally, we compute the contact and vertex probabilities ${\langle\bm{1}_{uv}\rangle_i}$ and ${\langle\bm{1}_{u}\rangle_i}$.
Below, we first describe the construction of the Markov Chain and then provide details of these algorithms.

\subsection{Monte Carlo acceptance probabilities}

In order to calculate equilibrium properties of the contact-graph model, the underlying Markov Chain must obey detailed balance.
That is, the probability of making forward and reverse moves between two subgraphs $g$ and $g'$ must be equal.
To do so, we propose transitions between microstates (which obey the two rules given in \secref{sec:allowed_microstates}) with uniform probability and then correct for this bias by calculating the ratio of the generation probabilities between forward and backward moves, ${\alpha(g \rightarrow g') / \alpha(g' \rightarrow g)}$.

Assuming a single connected component (i.e., a single structured region) $c$, we implement moves that add or remove individual vertices.
The set of vertices that are adjacent to $c$ in the supergraph $G$ but are not in $\mathcal{V}(g)$ is denoted by ${\mathcal{A}(g,c)}$.
We choose one vertex $u$ from ${\mathcal{A}(g,c)}$ with uniform probability and form all edges ${(u,v) \in \mathcal{E}(G)}$ between $u$ and the existing vertices ${v \in \mathcal{V}(g,c)}$.
With the addition of these edges, we denote the new graph as $g'$ and the updated connected component as $c'$.

For the reverse move, we must avoid breaking $c'$ into two or more disconnected subgraphs.
Consequently, we must be careful not to remove any vertex that is an articulation point of $c'$.
The set of such points is denoted by ${\mathcal{B}(g'\!,c')}$.
We therefore select one vertex with uniform probability from the set ${\mathcal{V}(g'\!,c') \setminus \mathcal{B}(g'\!,c')}$.
For this move, we only consider connected components that are larger than a dyad.
The ratio of the forward to reverse generation probabilities is
\begin{equation}
  \frac{\alpha_{\text{N}+}(g,c \rightarrow g'\!,c')}{\alpha_{\text{N}-}(g'\!,c' \rightarrow g,c)} = \frac{|\mathcal{A}(g,c)|}{\bm{1}\big[|\mathcal{V}(g'\!,c')| > 2\big]^{\phantom{\circ}}\!\! |\mathcal{V}(g'\!,c') \setminus \mathcal{B}(g'\!,c')|}.
\end{equation}

To ensure ergodicity and to improve sampling efficiency, we implement a super-detailed balance sampling scheme~\cite{frenkel2001understanding} for vertex additions and removals.
If a move ${g,c \rightarrow g',c'}$ results in a subgraph that violates rule 2 in \secref{sec:allowed_microstates}, we immediately attempt another move of the same type, starting from the new subgraph ${g'}$ using the updated connected component ${c'}$.
This process is repeated until the resulting subgraph, ${g^{(n)}}$, satisfies rule 2.
The total probability of following this path from $g$ to ${g^{(n)}}$ is the product of the generation probabilities at each step, ${\alpha(g \rightarrow g^{(1)}) \times \alpha(g^{(1)} \rightarrow g^{(2)}) \times \cdots \times \alpha(g^{(n-1)} \rightarrow g^{(n)})}$.
The ratio of generation probabilities depends on the total probability of following this forward path and the total probability of traversing the path in reverse, following precisely the same sequence of steps:
\begin{equation}
  \frac{\alpha_{\text{N}+}^{(n)}}{\alpha_{\text{N}-}^{(n)}} =
  \frac{\prod_{i = 1}^{n} \alpha_{\text{N}+}(g^{(i-1)},c^{(i-1)} \rightarrow g^{(i)}\!,c^{(i)})}{\prod_{i = 0}^{n-1} \alpha_{\text{N}-}(g^{(n-i)}\!,c^{(n-i)} \rightarrow g^{(n-i-1)},c^{(n-i-1)})},
\end{equation}
where each step is indexed by $i$ and ${g^{(0)} \equiv g}$.
If at any step on the forward move we find that ${|\mathcal{A}(g^{(i)},c^{(i)})| = 0}$, then the entire move is rejected.
In order to obey detailed balance, vertex additions and removals are attempted with equal probability at every Monte Carlo step.

\subsection{Wang--Landau sampling}

Wang--Landau sampling~\cite{wang2001efficient} provides an efficient algorithm for calculating the free-energy difference between two disjoint sets of microstates.
Here we implement the variant of this algorithm described in \refcite{belardinelli2007wang}.
In essence, the Wang--Landau algorithm calculates an equilibrium free-energy landscape stochastically by continually updating an estimate of the free energy, ${F_t}$, as the Monte Carlo calculation samples from the space of allowed subgraphs.
At every step, the underlying Monte Carlo algorithm uses ${F_t}$ to bias the acceptance probabilities of individual moves.

For these calculations, we use an order parameter to measure progress toward the completely folded microstate.
Excluding the effects of the backbone connectivity, which are entirely contained in ${\DSl(g)}$, the entropic contribution to the free energy in \eqref{eq:Fg} is proportional to
\begin{equation}
  X(g) \equiv \!\! \sum_{c \in \mathcal{C}(g)} \!\!\! \big[ |\mathcal{V}(g,c)| - 1 \big] = N(g) - C(g),
  \label{eq:X}
\end{equation}
where $N(g)$ is the total number of interacting residues and ${C(g) \equiv |\mathcal{C}(g)|}$ is the number of connected components of the microstate $g$.
Like the commonly used fraction of native contacts, $Q$~\cite{onuchic2004theory}, the order parameter $X$ characterizes the similarity between any given microstate and the native configuration.
However, $X$ is preferable for analyzing a discrete model, since it measures the degree of assembly of the independent monomers as opposed to the (likely correlated) interactions among them.
Since our calculations only consider the largest structured region, ${C(g) = 1}$ for all topological configurations except $\varnothing$, in which case ${C(g) = 0}$.

To perform free-energy calculations for a specific topological configuration~$i$, we first find the subgraph of $G$ that contains the maximum number of compatible contacts.
(We find the maximal subgraph containing all possible edges from all substructures in topological configuration~$i$, without including edges from substructures that are not represented in configuration~$i$.)
The free energy of this microstate serves as the reference state for the Wang--Landau calculation, ${F[i,\max_i(X)]}$.
We then apply the algorithm described in \refcite{belardinelli2007wang} using the following acceptance probabilities for proposed moves ${g \rightarrow g'}$:
\begin{eqnarray}
  \label{eq:pacc}
  p_{\text{acc}}(g \!\rightarrow\! g') &=& \min \!\Bigg\{\! 1, \, \frac{\alpha(g'\! \rightarrow\! g)}
  {\alpha(g\! \rightarrow\! g')} \, e^{-\big[ F(g') - F(g) \big]/\kB T} \\
  &\quad& \qquad\qquad\! \times\, e^{ \Big[F_t\big[i,X(g')\big] - F_t\big[i,X(g)\big]\Big] / \kB T} \Bigg\}.\;\,
  \nonumber
\end{eqnarray}
The Wang--Landau algorithm breaks detailed balance, since the bias changes as a function of the Monte Carlo `time,' $t$.
However, the amount by which ${F_t(i,X)}$ is updated between Monte Carlo moves is gradually decreased as the algorithm runs such that the estimated ${F_t(i,X)}$ converges to the equilibrium free-energy landscape.
The total free energy of each topological configuration is then ${F_i = -\kB T \ln \sum_X \exp[-F_t(i,X) / \kB T]}$.
For proteins with $\sim\!60$ residues, sufficiently converged results for all topological configurations can typically be obtained in a few minutes on a single processor.

\subsection{Calculation of ensemble averages}

Once the Wang--Landau sampling is complete, we use ${F_t(i,X)}$ as a biasing potential to accelerate the calculation of equilibrium averages via standard Metropolis Monte Carlo sampling.
If the free-energy differences between adjacent coarse-grained states have converged to within ${\sim\! 1~\kB T}$, then biased Metropolis Monte Carlo sampling will visit all coarse-grained states with roughly equal frequency.
This means that the Metropolis algorithm can provide a direct verification of the convergence of the Wang--Landau sampling.

We use Metropolis Monte Carlo sampling to compute the equilibrium contact probability, ${\langle \bm{1}_{uv} \rangle}$, and vertex probability, ${\langle \bm{1}_u \rangle}$, within each topological configuration~$i$.
We calculate the probability that the contact ${(u,v)}$ or the vertex $u$ appears in the set of visited microstates,
\begin{eqnarray}
  \langle \bm{1}_{uv} \rangle_i &\simeq& \frac{\sum_X\! \sum_{\{y\}_X}\! \bm{1}_{uv}(g_y) \, e^{-F(i,X)/\kB T}}{\sum_X\! \sum_{\{y\}_X}\! e^{-F(i,X)/\kB T}}, \\
  \langle \bm{1}_{u} \rangle_i &\simeq& \frac{\sum_X\! \sum_{\{y\}_X}\! \bm{1}_{u}(g_y) \, e^{-F(i,X)/\kB T}}{\sum_X\! \sum_{\{y\}_X}\! e^{-F(i,X)/\kB T}},
\end{eqnarray}
where $\bm{1}_{uv}(g)$ and $\bm{1}_u(g)$ indicate the presence of edge ${(u,v)}$ and vertex $u$, respectively, in microstate $g$, and ${\{y\}_X}$ is the set of all visited microstates with order parameter $X$.
The use of a biasing potential allows the Markov chain to explore the entire free-energy landscape rapidly without getting stuck for long intervals in local free-energy minima.
The fact that the underlying Markov chain obeys detailed balance ensures that the ensemble average within each coarse-grained state ${(i,X)}$ converges to its equilibrium value given a sufficient number of Monte Carlo steps, ${n_{\text{MC}}}$.
Typically, we choose ${n_{\text{MC}} \simeq 1000}$ per coarse-grained state ${(i,X)}$.

\section{Mean-field barrier calculations}

To compute the free-energy barrier between a pair of topological configurations $i$ and $j$, we assume that the initial configuration~$i$ is in local equilibrium.
Making a contact between the existing structure in configuration~$i$ and the new substructure $s$, which is part of configuration~$j$, necessarily requires the formation of a new loop in the polymer backbone; after this initial contact, folding can proceed in topological configuration~$j$ by making further native contacts at a much smaller entropic cost per contact.
The barrier calculation should therefore account for all the ways in which this initial contact between the structured region of configuration~$i$ and the new substructure $s$ can be made.
This calculation is carried out in a mean-field approximation, where the effective strength of an interaction between a residue from the new substructure $s$ and a residue $v$ in the existing structured region depends on the local equilibrium in configuration~$i$, ${\langle \bm{1}_v \rangle_i}$; this approximation is described below.
Fluctuations within configuration~$i$ are taken into account by Boltzmann-averaging this barrier calculation over all values of the order parameter $X$ in this configuration.

\begin{figure}
  \includegraphics{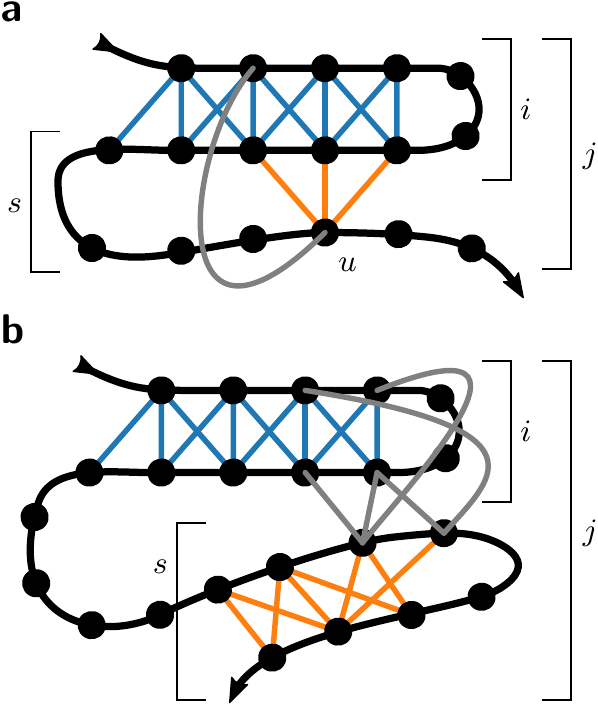}
  \caption{\textbf{Schematic of mean-field barrier calculations.}  (a) In the first mechanism, a single vertex $u$ is added to the existing structured region $i$ to form a new loop in the polymer backbone.  (b) In the second mechanism, a pre-assembled substructure $s$ makes contact with the existing structured region~$i$; in this case, substructure $s$ has no residues in common with configuration~$i$.  In both cases, after the formation of this initial contact, the polymer is in topological configuration~$j$.  See text for details.}
  \label{fig:barriers}
\end{figure}

The addition of a new substructure to the existing structured region in configuration~$i$ can occur by one of two mechanisms, depending on the way the substructures interact in topological configuration~$j$.
The first mechanism applies in cases where the contacts associated with the new substructure $s$ directly involve residues that are already present in topological configuration~$i$.
As a result, the first step in the assembly of the new substructure involves the addition of a residue $u$ that participates in substructure $s$ but is not part of the existing structured region $i$ (see \figref{fig:barriers}a).
Assuming that the value of the order parameter for the existing structure is $X$, the mean-field free energy of this configuration depends on the loss of conformation entropy due to bringing a residue $u$ into contact with structured region in configuration~$i$, ${\langle \Delta S_u \rangle_{i,X}}$, as well as the mean-field energies of all native contacts between $u$ and residues in region $i$,
\begin{eqnarray}
  \frac{\Delta F^\dagger_{i,X \rightarrow j}}{\kB T} &=&
  -\ln \sum_u \exp \left\langle \frac{\Delta S_u}{\kB} \right\rangle_{\!i,X}
  \label{eq:barrier1} \\
  &\,&\qquad \times \left\{ \exp \!\left[ -\!\!\!\sum_{v \in \mathcal{V}(i)} \!\!\!\left(\frac{\epsilon_{uv}}{\kB T}\right) \langle \bm{1}_v \rangle_{i,X} \right] - 1 \!\right\}, \nonumber
\end{eqnarray}
where $\mathcal{V}(i)$ indicates the set of residues that contribute to configuration~$i$.
The first sum in \eqref{eq:barrier1} runs over all residues ${\{u\}}$ that participate in one of the contacts comprising substructure~$s$ and are not in the set $\mathcal{V}(i)$.

The second mechanism applies in cases where the new and existing substructures do not have any residues in common (see \figref{fig:barriers}b).
Instead, these substructures interact in the native state via edges that are not part of any substructure (i.e., gray edges in \figref{fig:barriers}b).
To calculate the barrier in this case, we assume that both the initial configuration~$i$ and the new substructure~$s$ are in local equilibrium.
In the mean-field approximation, the free energy of all microstates in which the substructure~$s$ makes contact with the locally equilibrated structured region in configuration~$i$, assuming that the value of the order parameter for the existing structure is~$X$, is
\begin{eqnarray}
  \frac{\Delta F^\dagger_{i,X \rightarrow j}}{\kB T} &=&
  F_s - \left\langle \frac{\Delta S_s}{\kB} \right\rangle_{\!i,X} \\
  &\,& \!\!\!- \ln \!\left\{ \!\exp \!\left[ -\!\!\!\sum_{\substack{u \in \mathcal{V}(s)\\v \in \mathcal{V}(i)}} \!\!\!\langle \bm{1}_u \rangle_{s} \left(\frac{\epsilon_{uv}}{\kB T}\right) \langle \bm{1}_v \rangle_{i,X} \!\right]\! - 1 \!\right\}\!, \nonumber
\end{eqnarray}
where $F_s$ is the free energy of the isolated substructure~$s$ and ${\langle \Delta S_s \rangle_{i,X}}$ is the entropic penalty due to bringing~$s$ into contact with the structured region in configuration~$i$.

We compute the apparent barrier between configurations $i$ and $j$ by summing over all values of the order parameter~$X$,
\begin{equation}
  \frac{\Delta F^\dagger_{i \rightarrow j}}{\kB T} \!= \!-\ln \sum_X \exp \!\left[ \frac{-\Delta F^\dagger_{i,X \rightarrow j} - \left(F_{i,X} - F_i\right)}{\kB T} \right]\!.\!
\end{equation}
The term ${(F_{i,X} - F_i)}$ accounts for the free-energy difference between microstates at a specific value of the order parameter and the total free energy of topological configuration~$i$, $F_i$.
To obey detailed balance, the barrier for the reverse transition is ${\Delta F^\dagger_{j \rightarrow i} = \Delta F^\dagger_{i \rightarrow j} - (F_j - F_i)}$.

\section{Transition-path theory}
\label{sec:tpt}

Given the continuous-time Markov chain specified by the rate matrix in \eqref{eq:k}, we can use transition-path theory~\cite{metzner2009transition} to calculate properties of the ensemble of folding trajectories.
The stationary distribution of the Markov chain, ${\pi(i,X)}$, is equivalent to the Boltzmann distribution, ${\pi_i = \exp(-F_i / \kB T) / \sum_j \exp(-F_i / \kB T)}$.
All folding transition paths originate in the unfolded configuration, ${A = \varnothing}$, and terminate in the configuration with the maximum number of substructures, $B$.
Here we reproduce a number of equations from \refcite{metzner2009transition} for completeness.

First, we calculate ${\pfold(i)}$, the equilibrium probability that a dynamical trajectory will reach configuration~$B$, starting from configuration~$i$, before returning to configuration~$A$.
By definition, $\pfold$ is equal to zero and one in configurations $A$ and $B$, respectively.
Using the rate matrix $k_{ij}$, $\pfold$ is computed for all intermediate configurations by solving the linear system
\begin{equation}
  \label{eq:pfold}
  \sum_j \! k_{ij} \pfold(j) = 0 \quad \forall i \in ( A \cup B )^c,
\end{equation}
where ${( A \cup B )^c}$ indicates all configurations that are neither $A$ nor $B$.
The reactive flux through every transition ${i \rightarrow j}$ is
\begin{equation}
  f(i \rightarrow j) =
  \begin{cases}
    \pi_i \big[1 - \pfold(i)\big] k_{ij} \pfold(j)  & \text{if } i \ne j, \\
    0 & \text{if } i = j.
  \end{cases}
\end{equation}
The net reactive flux through the transition ${i \rightarrow j}$ is ${f^+_{ij} \equiv \max(f_{ij} - f_{ji}, 0)}$.
From this calculation, we can determine the overall folding rate,
\begin{equation}
  \kfold = \frac{\sum_{j \ne A} f^+_{Aj}}{\pi_A}.
\end{equation}
In the two-state approximation, the apparent free-energy barrier between configurations $A$ and $B$ is ${\Delta F^{\dagger}_{AB} = -\ln\left(2\kfold\right)}$.
Lastly, the fraction of time spent in configuration~$i$ in the transition-path ensemble is
\begin{equation}
  p_{AB}(i) = \pi_i \pfold(i) [1 - \pfold(i)].
\end{equation}

\section{Theoretical $\phi$ and $\psi$-value calculations}
\label{sec:phi_psi}

Theoretical $\phi$ and $\psi$-values are calculated as described in \eqsref{eq:psi} and (\ref{eq:phi}).
For the rate calculation, $\kfold$, the unfolded, $A$, and folded, $B$, configurations are chosen as described in \secref{sec:tpt}.
The inverse temperature ${(\kB T)^{-1}}$ is chosen to equate the free-energies of the folded ensemble, which includes contributions from all native contacts, and the unfolded ensemble; we take the unfolded ensemble to include both the random coil configuration, $\varnothing$, and all individual substructures in isolation, ${F_{\text{unfolded}} = -\kB T \ln \left[1 + \sum_s \exp(-F_s/\kB T)\right]}$, where the index $s$ runs over all substructures.
(Exceptions are made for proteins 1rnb and 2vil, where, due to the stability of partially folded intermediate configurations, the free-energy differences between the native and unfolded ensembles are set to $-2.5$ and $-1~\kB T$, respectively.  These choices ensure that the native states are globally stable.)

\begin{figure}
  \includegraphics{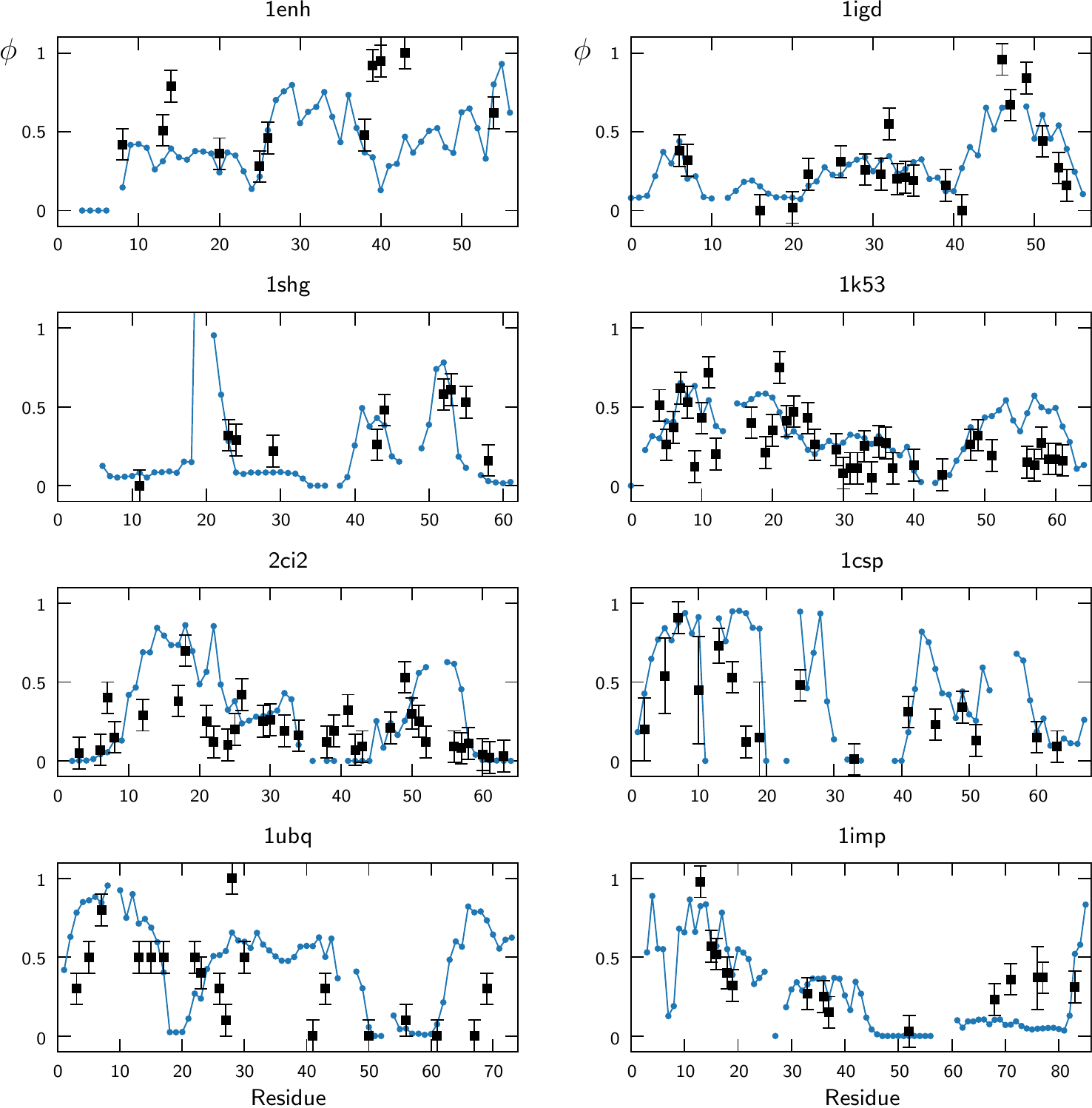}
  \caption{\textbf{Comparison of predicted and experimental $\phi$-values.}  Predictions are indicated by blue circles, and experimental points are shown as black squares.  Experimental errors are assumed to be $0.1$ unless otherwise indicated; see \tabsref{tab:phi1} and \ref{tab:phi2} for a list of data points.}
  \label{fig:phi1}
\end{figure}

\begin{figure}
  \includegraphics{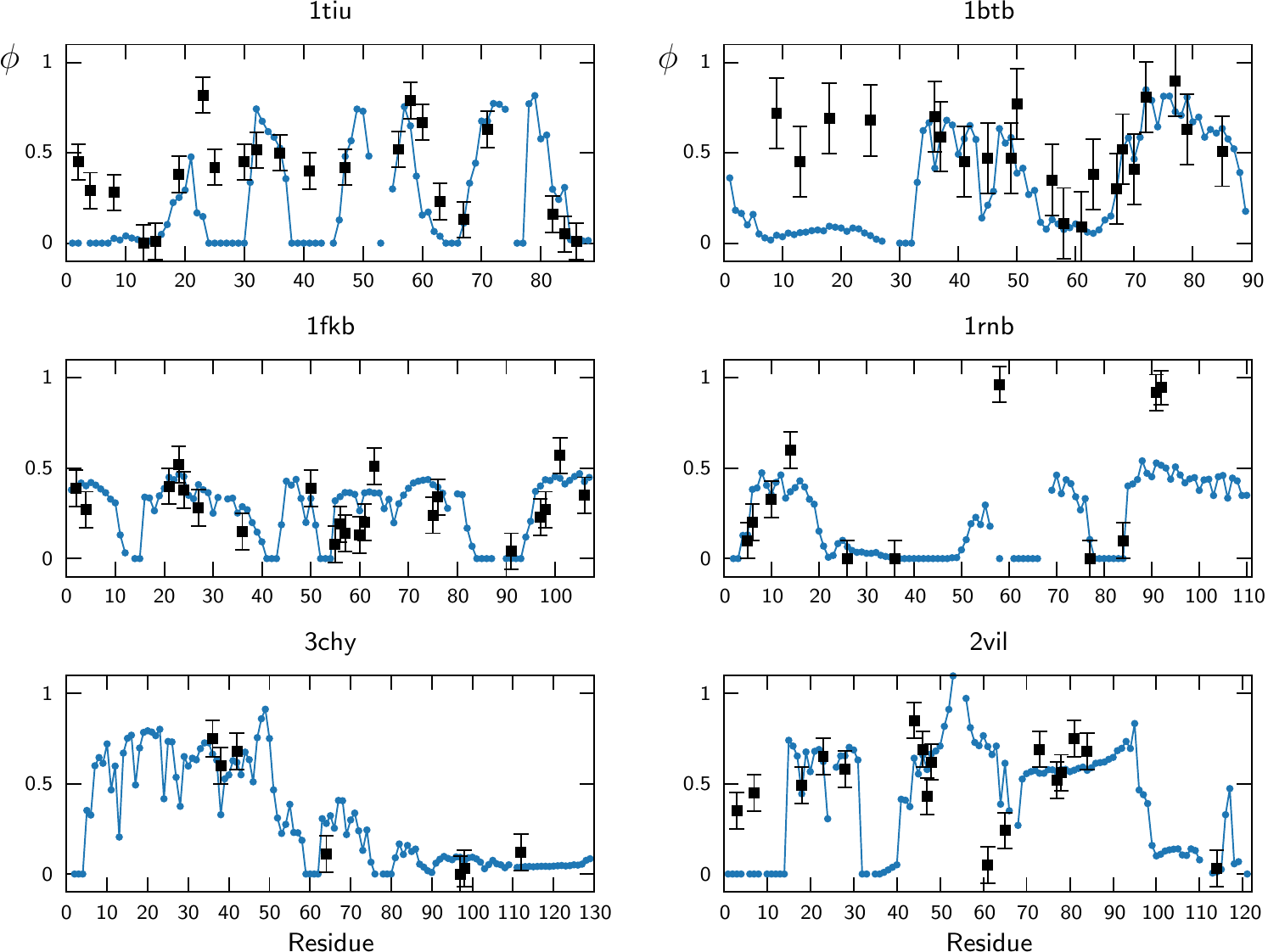}
  \caption{\textbf{Comparison of predicted and experimental $\phi$-values (continued).}  Predictions are indicated by blue circles, and experimental points are shown as black squares.  Experimental errors are assumed to be $0.1$ unless otherwise indicated; see \tabsref{tab:phi1} and \ref{tab:phi2} for a list of data points.}
  \label{fig:phi2}
\end{figure}

\begin{figure}
  \includegraphics{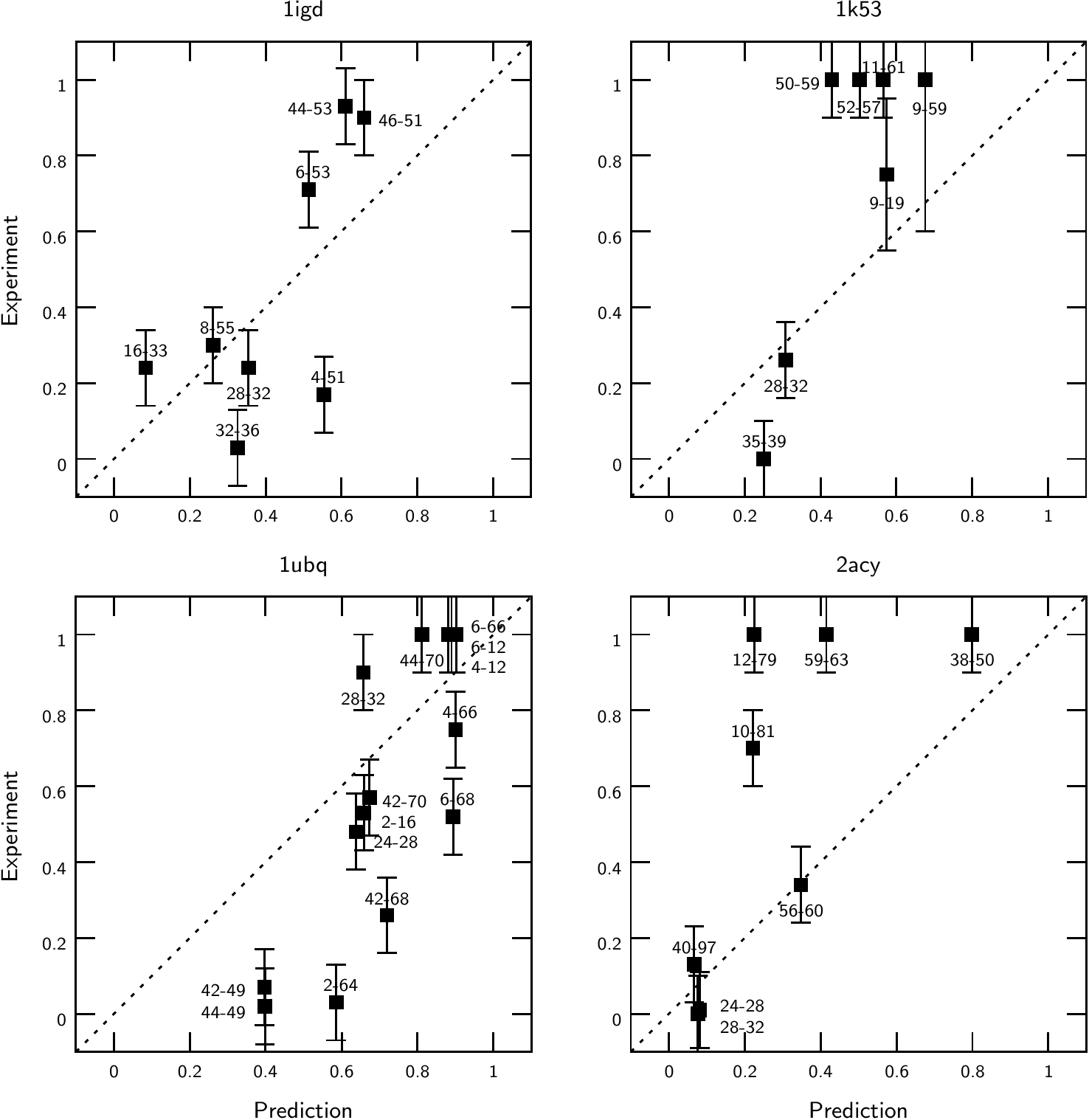}
  \caption{\textbf{Scatter plots for $\psi$-value comparisons.}  For each point, the pair of mutated residues is indicated.  Experimental errors are assumed to be $0.1$ unless otherwise indicated; see \tabref{tab:psi} for a list of data points.}
  \label{fig:psi}
\end{figure}

The mutations considered in our comparisons with experimental measurements are listed in \tabsref{tab:phi1}--\ref{tab:psi} and shown in \figsref{fig:phi1}--\ref{fig:psi}.
Unless otherwise noted, we assume that the experimental errors on $\phi$ and $\psi$-values are ${\pm 0.1}$.
We leave $\phi$-values that are less than ${-0.1}$ or greater than ${1.1}$ out of comparisons with the theoretical predictions.
($\phi$-values in the range $[-0.1:0]$ or $[1:1.1]$ are set to 0 or 1, respectively.)
For $\psi$-value comparisons, we set values greater than 1 to unity.

\section{Analysis of atomistic Molecular Dynamics simulations}

For the analysis of atomistic simulation data, we adopt a history-dependent native-contact definition~\cite{best2013native}: a contact is formed when heavy atoms from two residues pass within 3.5~\AA~of one another and broken when all heavy atoms of the same residues move farther than 5.5~\AA~apart.
To reduce the contribution of transient fluctuations further, we disregarded contacts lasting less than 5~ns; changing this threshold by ${\pm5}$~ns does not meaningfully affect the results of the subsequent calculations.
Native contacts were defined on the basis of the crystal structure as described in \secref{sec:theory} for direct comparison with the theoretical results.
We determined the largest structured region at every 1-ns time step by decomposing the graph of native contacts into connected components.
We then calculated a one-dimensional free-energy landscape as a function of the number of native contacts using all time steps from the available trajectories.
Folding transition paths are defined as the portions of the trajectories that transit from the free-energy minimum of the unfolded ensemble on this landscape to the free-energy minimum of the folded ensemble without returning to the free-energy minimum of the unfolded ensemble.
Unfolding transition paths are defined analogously, starting from the free-energy minimum of the folded ensemble.

For the configuration lifetime calculations shown in \figref{fig:assumptions}, we identified all substructures with at least 6 contacts present in the largest structured region.
We verified that every such substructure is completely contained within the largest structured region, i.e., no contacts from a substructure that forms part of the largest structured region are found outside of this region, in more than 99.8\% of all time steps.
For the commitment calculations shown in \figref{fig:commit}, we used the stricter criterion for substructure formation described in \secref{sec:results} of the main text.

We calculated $\phi$-values from the simulated transition paths using the method described in~\refcite{best2016microscopic},
\begin{eqnarray}
  \psi_{uv}^{\text{simulation}} &\simeq& p(\bm{1}_{uv}|\text{TP}), \\
  \phi_u^{\text{simulation}} &=& \sum_v \psi_{uv}^{\text{simulation}} / d_u,
\end{eqnarray}
where ${p(\bm{1}_{uv}|\text{TP})}$ is the probability of observing a native contact between residues $u$ and $v$ at any time step in the transition-path ensemble and $d_u$ is the number of native contacts formed by residue $u$.
We estimated the variability in the predicted $\phi$ and $\psi$-values across the observed transition paths by performing bootstrapping simulations in which the 10 observed transition paths were sampled with replacement; the standard deviation of $\phi_u^{\text{simulation}}$ estimated in this way is shown in \figref{fig:phi_psi_comparison}b.
The calculations shown in \figref{fig:phi_psi_comparison}b are slightly different from the results presented in \refcite{best2016microscopic} because our definitions of native contacts are not identical.
Because $\phi_u^{\text{simulation}}$ is calculated directly from $\psi_{uv}^{\text{simulation}}$, we obtain the same correlation coefficient with the theoretical predictions for both sets of values.

\onecolumngrid
\clearpage

\begin{figure}[h]
  \includegraphics{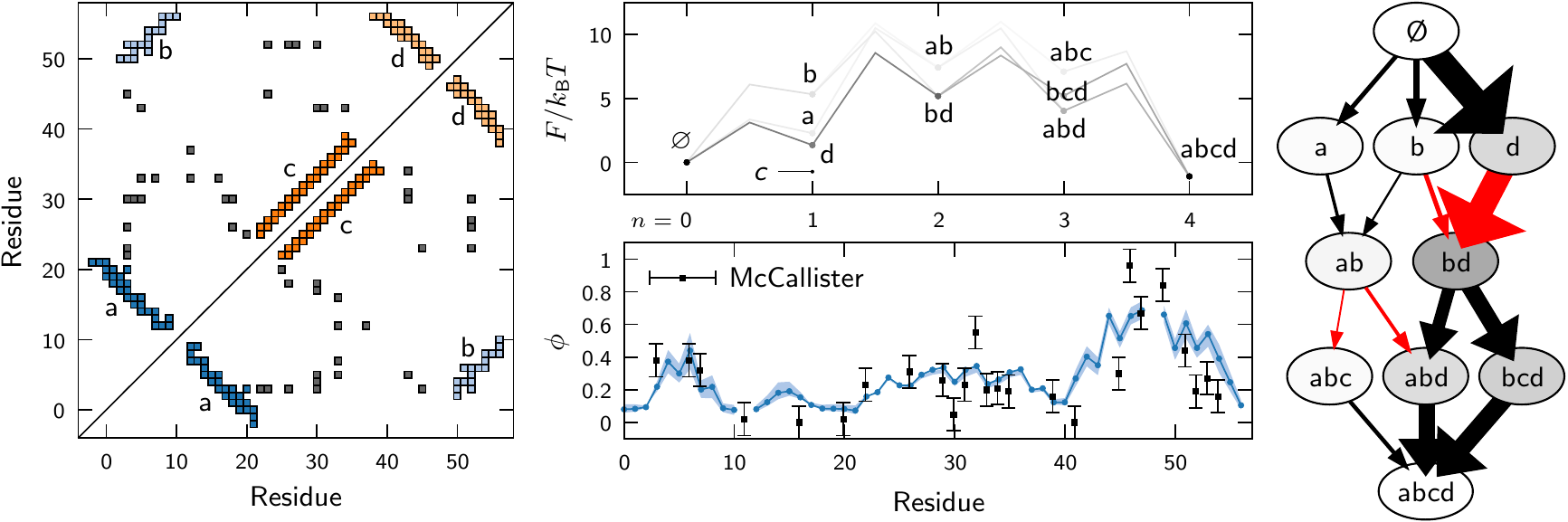}
  \caption{\textbf{Predicted folding landscape for protein G (1igd) and comparison with experimental $\phi$-values~\cite{mccallister2000critical}.}  The configuration \textsf{abcd} is the native ensemble in this case, because all residues contribute the one of the four substructures.  The free-energy landscape and folding network are drawn as in \figsref{fig:landscapes}c and \ref{fig:network}a, respectively.}
  \label{fig:1igd}
\end{figure}

\begin{figure}[h]
  \includegraphics{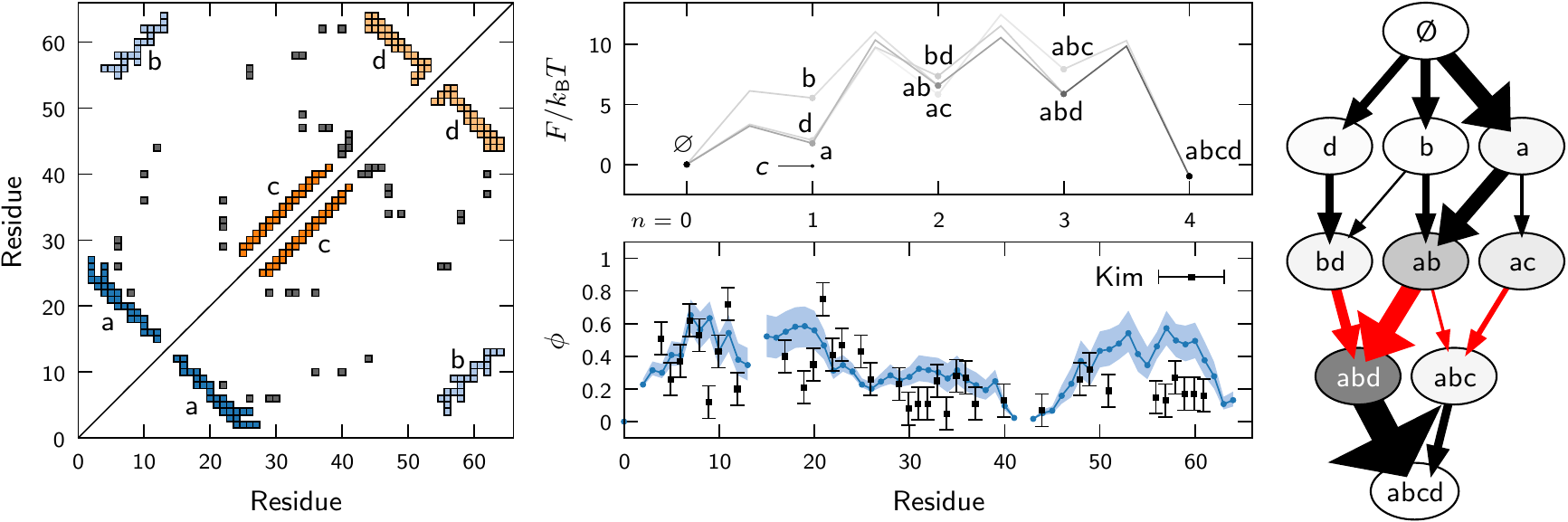}
  \caption{\textbf{Predicted folding landscape for protein L (1k53) and comparison with experimental $\phi$-values~\cite{kim2000breakdown}.}  The configuration \textsf{abcd} is the native ensemble in this case, because all residues contribute the one of the four substructures.  The free-energy landscape and folding network are drawn as in \figsref{fig:landscapes}c and \ref{fig:network}a, respectively.}
  \label{fig:1k53}
\end{figure}

\begin{figure}[h]
  \includegraphics{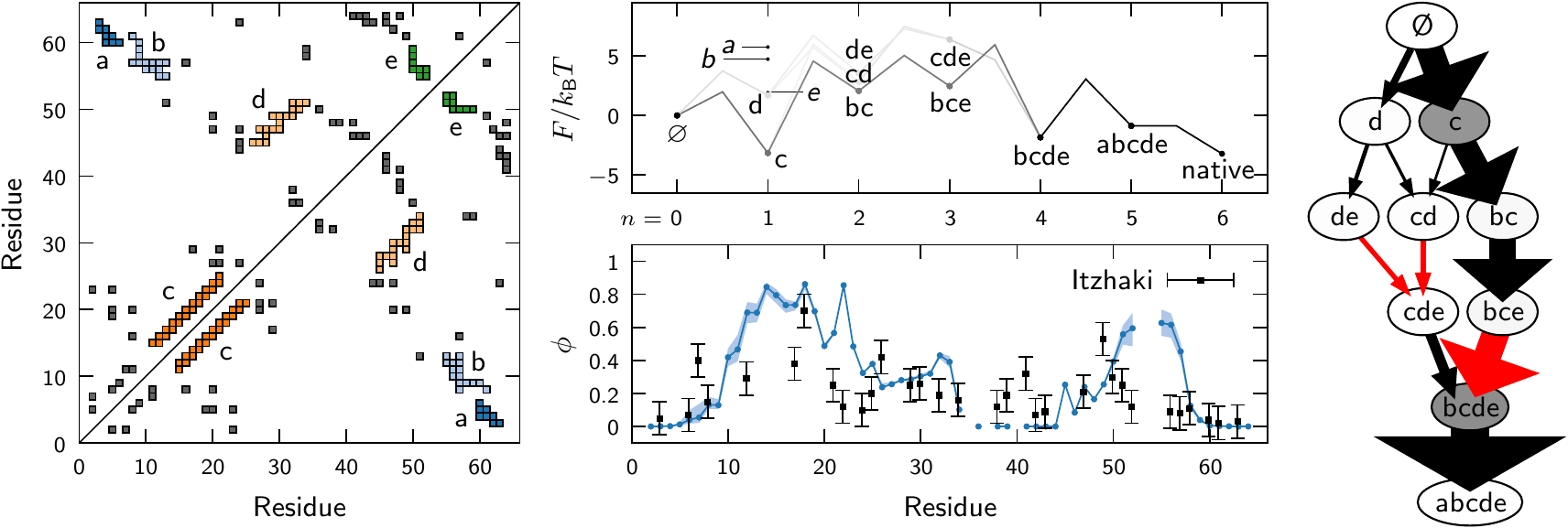}
  \caption{\textbf{Predicted folding landscape for chymotrypsin inhibitor 2 (2ci2) and comparison with experimental $\phi$-values~\cite{itzhaki1995structure}.}  The free-energy landscape and folding network are drawn as in \figsref{fig:landscapes}c and \ref{fig:network}a, respectively.}
  \label{fig:2ci2}
\end{figure}

\begin{figure}[h]
  \includegraphics{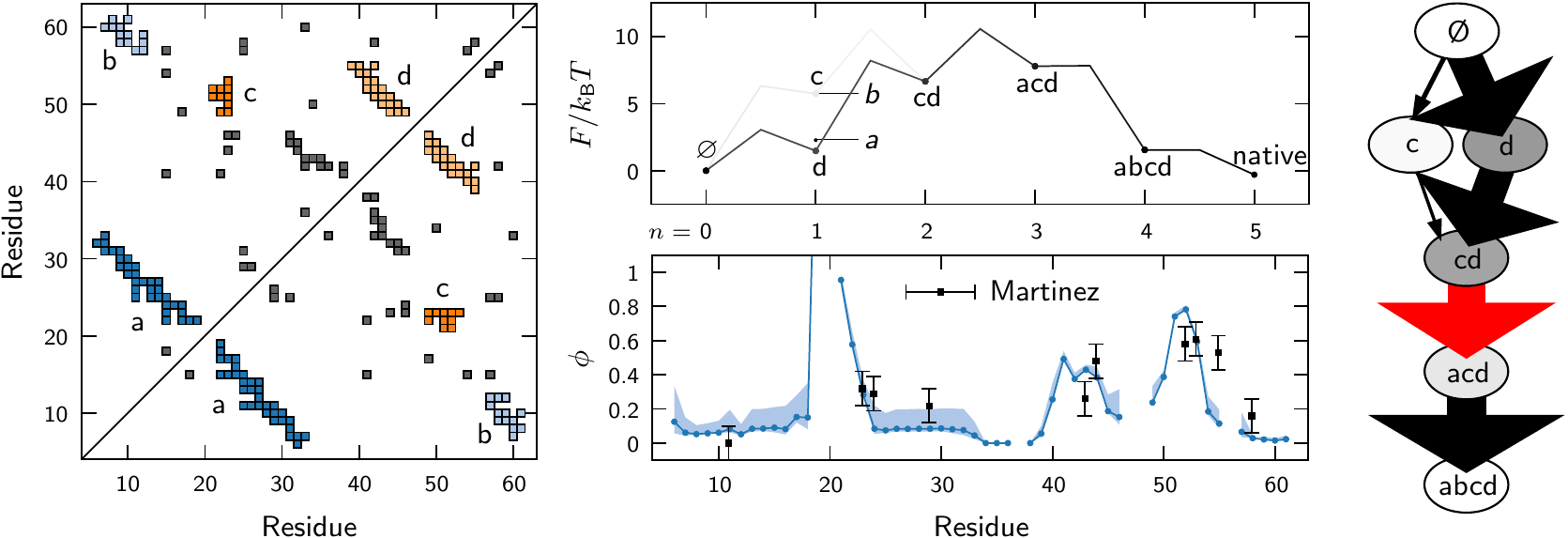}
  \caption{\textbf{Predicted folding landscape for the Src SH3 domain (1shg) and comparison with experimental $\phi$-values~\cite{martinez1999folding}.}  The free-energy landscape and folding network are drawn as in \figsref{fig:landscapes}c and \ref{fig:network}a, respectively.}
  \label{fig:1shg}
\end{figure}

\begin{figure}[h]
  \includegraphics{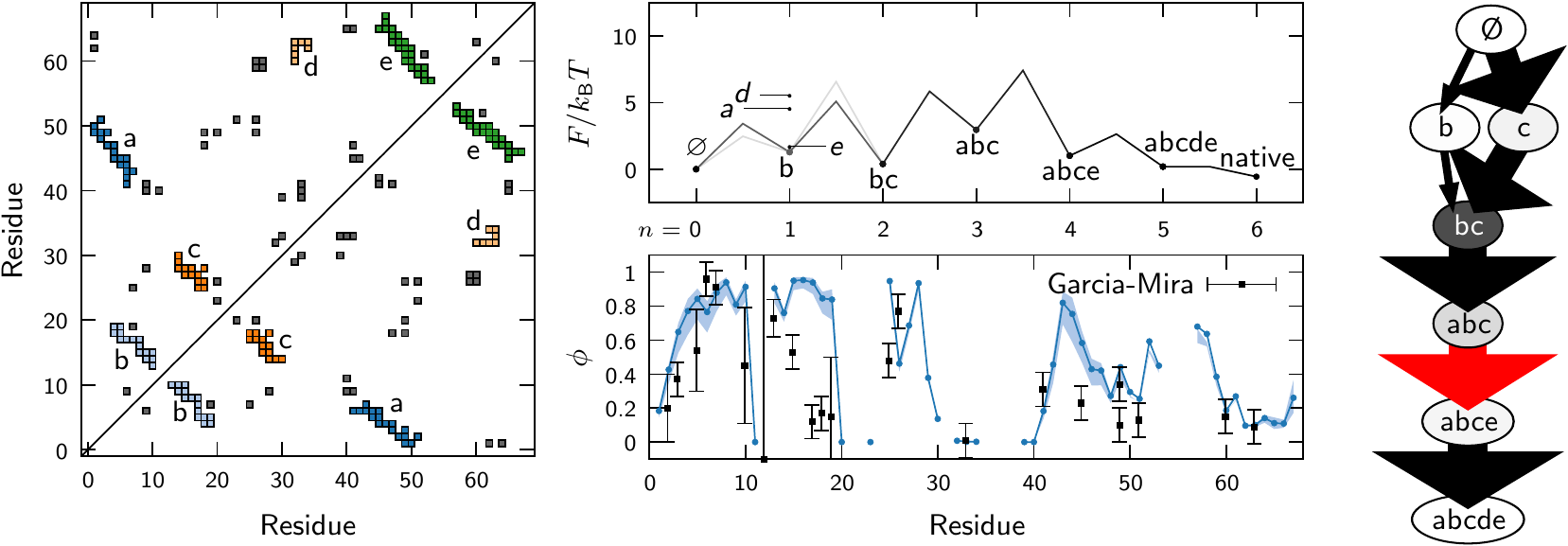}
  \caption{\textbf{Predicted folding landscape for the cold-shock protein (1csp) and comparison with experimental $\phi$-values~\cite{garcia2004folding}.}  The free-energy landscape and folding network are drawn as in \figsref{fig:landscapes}c and \ref{fig:network}a, respectively.}
  \label{fig:1csp}
\end{figure}

\begin{table*}
  \begin{minipage}[t]{4.25cm}
    \begin{tabular}{lll}
      \textbf{1enh} & $\phi_{\text{expt}}$ & $\phi_{\text{pred}}$ \\
      \hline
      F8A & 0.42$\pm$0.1 & 0.15 \\
      L13A & 0.51$\pm$0.1 & 0.31 \\
      A14G & 0.79$\pm$0.1 & 0.39 \\
      F20A & 0.36$\pm$0.1 & 0.24 \\
      Y25G & 0.28$\pm$0.1 & 0.22 \\
      L26A & 0.46$\pm$0.1 & 0.51 \\
      L38A & 0.48$\pm$0.1 & 0.37 \\
      G39A & 0.92$\pm$0.1 & 0.34 \\
      L40A & 0.95$\pm$0.1 & 0.13 \\
      A43G & 1.00$\pm$0.1 & 0.47 \\
      A54G & 0.62$\pm$0.1 & 0.80 \\
      \rule{0pt}{6ex}
      \textbf{1igd} & $\phi_{\text{expt}}$ & $\phi_{\text{pred}}$ \\
      \hline
      I6A & 0.38$\pm$0.1 & 0.44 \\
      L7A & 0.32$\pm$0.1 & 0.20 \\
      T16A & 0.00$\pm$0.1 & 0.15 \\
      A20G & 0.02$\pm$0.1 & 0.08 \\
      D22A & 0.23$\pm$0.1 & 0.16 \\
      A26G & 0.31$\pm$0.1 & 0.23 \\
      V29A & 0.26$\pm$0.1 & 0.34 \\
      K31G & 0.23$\pm$0.1 & 0.32 \\
      Q32G & 0.55$\pm$0.1 & 0.35 \\
      Y33A & 0.20$\pm$0.1 & 0.24 \\
      A34G & 0.21$\pm$0.1 & 0.26 \\
      N35G & 0.19$\pm$0.1 & 0.31 \\
      V39A & 0.16$\pm$0.1 & 0.12 \\
      G41A & 0.00$\pm$0.1 & 0.27 \\
      D46A & 0.96$\pm$0.1 & 0.65 \\
      D47A & 0.67$\pm$0.1 & 0.69 \\
      T49A & 0.84$\pm$0.1 & 0.66 \\
      T51A & 0.44$\pm$0.1 & 0.61 \\
      T53A & 0.27$\pm$0.1 & 0.54 \\
      V54A & 0.16$\pm$0.1 & 0.39 \\
      \rule{0pt}{6ex}
      \textbf{1shg} & $\phi_{\text{expt}}$ & $\phi_{\text{pred}}$ \\
      \hline
      A11G & 0.00$\pm$0.1 & 0.08 \\
      V23A & 0.32$\pm$0.1 & 0.28 \\
      T24A & 0.29$\pm$0.1 & 0.08 \\
      D29A & 0.22$\pm$0.1 & 0.08 \\
      K43A & 0.26$\pm$0.1 & 0.43 \\
      V44A & 0.48$\pm$0.1 & 0.38 \\
      F52A & 0.58$\pm$0.1 & 0.78 \\
      V53A & 0.61$\pm$0.1 & 0.59 \\
      A55G & 0.53$\pm$0.1 & 0.11 \\
      V58A & 0.16$\pm$0.1 & 0.03
    \end{tabular}
  \end{minipage}
  \begin{minipage}[t]{4.25cm}
    \begin{tabular}{lll}
      \textbf{1k53} & $\phi_{\text{expt}}$ & $\phi_{\text{pred}}$ \\
      \hline
      V4A & 0.51$\pm$0.1 & 0.30 \\
      T5A & 0.26$\pm$0.1 & 0.41 \\
      I6A & 0.37$\pm$0.1 & 0.41 \\
      K7A & 0.62$\pm$0.1 & 0.65 \\
      A8G & 0.53$\pm$0.1 & 0.56 \\
      N9A & 0.12$\pm$0.1 & 0.63 \\
      L10A & 0.43$\pm$0.1 & 0.43 \\
      I11A & 0.72$\pm$0.1 & 0.54 \\
      F12A & 0.20$\pm$0.1 & 0.38 \\
      T17A & 0.40$\pm$0.1 & 0.55 \\
      T19A & 0.21$\pm$0.1 & 0.58 \\
      A20G & 0.35$\pm$0.1 & 0.56 \\
      E21A & 0.75$\pm$0.1 & 0.47 \\
      F22A & 0.41$\pm$0.1 & 0.31 \\
      K23A & 0.47$\pm$0.1 & 0.35 \\
      T25A & 0.43$\pm$0.1 & 0.23 \\
      F26G & 0.26$\pm$0.1 & 0.20 \\
      A29G & 0.23$\pm$0.1 & 0.25 \\
      T30A & 0.08$\pm$0.1 & 0.27 \\
      S31G & 0.11$\pm$0.1 & 0.32 \\
      E32G & 0.11$\pm$0.1 & 0.32 \\
      A33G & 0.25$\pm$0.1 & 0.30 \\
      Y34A & 0.05$\pm$0.1 & 0.26 \\
      A35G & 0.28$\pm$0.1 & 0.32 \\
      Y36A & 0.27$\pm$0.1 & 0.25 \\
      A37G & 0.11$\pm$0.1 & 0.22 \\
      L40A & 0.13$\pm$0.1 & 0.10 \\
      N44A & 0.07$\pm$0.1 & 0.05 \\
      T48A & 0.26$\pm$0.1 & 0.37 \\
      V49A & 0.32$\pm$0.1 & 0.32 \\
      V51A & 0.19$\pm$0.1 & 0.44 \\
      Y56A & 0.15$\pm$0.1 & 0.46 \\
      T57A & 0.13$\pm$0.1 & 0.57 \\
      L58A & 0.27$\pm$0.1 & 0.50 \\
      N59A & 0.17$\pm$0.1 & 0.47 \\
      I60A & 0.17$\pm$0.1 & 0.49 \\
      K61A & 0.16$\pm$0.1 & 0.38 \\
      \rule{0pt}{6ex}
      \textbf{2ci2} & $\phi_{\text{expt}}$ & $\phi_{\text{pred}}$ \\
      \hline
      T3G & 0.05$\pm$0.1 & 0.00 \\
      P6A & 0.07$\pm$0.1 & 0.04 \\
      E7A & 0.40$\pm$0.1 & 0.05 \\
      L8A & 0.15$\pm$0.1 & 0.13 \\
      S12G & 0.29$\pm$0.1 & 0.69 \\
      K17G & 0.38$\pm$0.1 & 0.74
    \end{tabular}
  \end{minipage}
  \begin{minipage}[t]{4.25cm}
    \begin{tabular}{lll}
      K18G & 0.70$\pm$0.1 & 0.86 \\
      L21A & 0.25$\pm$0.1 & 0.57 \\
      Q22G & 0.12$\pm$0.1 & 0.85 \\
      K24G & 0.10$\pm$0.1 & 0.32 \\
      P25A & 0.20$\pm$0.1 & 0.38 \\
      E26A & 0.42$\pm$0.1 & 0.24 \\
      I29A & 0.25$\pm$0.1 & 0.29 \\
      I30G & 0.26$\pm$0.1 & 0.30 \\
      L32A & 0.19$\pm$0.1 & 0.43 \\
      V34G & 0.16$\pm$0.1 & 0.10 \\
      V38A & 0.12$\pm$0.1 & 0.00 \\
      T39A & 0.19$\pm$0.1 & 0.00 \\
      E41A & 0.32$\pm$0.1 & 0.00 \\
      Y42G & 0.07$\pm$0.1 & 0.00 \\
      R43A & 0.09$\pm$0.1 & 0.00 \\
      V47A & 0.21$\pm$0.1 & 0.24 \\
      L49A & 0.53$\pm$0.1 & 0.26 \\
      F50A & 0.30$\pm$0.1 & 0.39 \\
      V51A & 0.25$\pm$0.1 & 0.56 \\
      D52A & 0.12$\pm$0.1 & 0.59 \\
      N56A & 0.09$\pm$0.1 & 0.62 \\
      I57A & 0.08$\pm$0.1 & 0.45 \\
      A58G & 0.11$\pm$0.1 & 0.13 \\
      V60G & 0.04$\pm$0.1 & 0.00 \\
      P61A & 0.02$\pm$0.1 & 0.00 \\
      V63G & 0.03$\pm$0.1 & 0.00 \\
      \rule{0pt}{6ex}
      \textbf{1csp} & $\phi_{\text{expt}}$ & $\phi_{\text{pred}}$ \\
      \hline
      L2A & 0.20$\pm$0.2 & 0.43 \\
      K5A & 0.54$\pm$0.24 & 0.84 \\
      K7A & 0.91$\pm$0.1 & 0.88 \\
      N10A & 0.45$\pm$0.34 & 0.91 \\
      K13A & 0.73$\pm$0.11 & 0.90 \\
      F15A & 0.53$\pm$0.1 & 0.95 \\
      F17A & 0.12$\pm$0.1 & 0.94 \\
      E19A & 0.15$\pm$0.35 & 0.84 \\
      D25A & 0.48$\pm$0.1 & 0.95 \\
      I33A & 0.01$\pm$0.1 & 0.00 \\
      L41A & 0.31$\pm$0.1 & 0.18 \\
      Q45A & 0.23$\pm$0.1 & 0.58 \\
      F49A & 0.34$\pm$0.1 & 0.44 \\
      I51A & 0.13$\pm$0.1 & 0.26 \\
      A60G & 0.15$\pm$0.1 & 0.19 \\
      V63A & 0.09$\pm$0.1 & 0.10
    \end{tabular}
  \end{minipage}
  \begin{minipage}[t]{4.25cm}
    \begin{tabular}{lll}
      \textbf{1ubq} & $\phi_{\text{expt}}$ & $\phi_{\text{pred}}$ \\
      \hline
      I3A & 0.30$\pm$0.1 & 0.78 \\
      V5A & 0.50$\pm$0.1 & 0.86 \\
      T7A & 0.80$\pm$0.1 & 0.85 \\
      I13A & 0.50$\pm$0.1 & 0.71 \\
      L15A & 0.50$\pm$0.1 & 0.69 \\
      V17A & 0.50$\pm$0.1 & 0.40 \\
      T22A & 0.50$\pm$0.1 & 0.27 \\
      I23A & 0.40$\pm$0.1 & 0.24 \\
      V26A & 0.30$\pm$0.1 & 0.51 \\
      L27A & 0.10$\pm$0.1 & 0.54 \\
      A28G & 1.00$\pm$0.1 & 0.66 \\
      I30A & 0.50$\pm$0.1 & 0.60 \\
      Q41A & 0.00$\pm$0.1 & 0.57 \\
      L43A & 0.30$\pm$0.1 & 0.50 \\
      L50A & 0.00$\pm$0.1 & 0.06 \\
      L56A & 0.10$\pm$0.1 & 0.05 \\
      I61A & 0.00$\pm$0.1 & 0.08 \\
      L67A & 0.00$\pm$0.1 & 0.79 \\
      L69A & 0.30$\pm$0.1 & 0.73 \\
      \rule{0pt}{6ex}
      \textbf{1imp} & $\phi_{\text{expt}}$ & $\phi_{\text{pred}}$ \\
      \hline
      A13G & 0.98$\pm$0.1 & 0.83 \\
      F15A & 0.57$\pm$0.1 & 0.58 \\
      L16A & 0.52$\pm$0.1 & 0.57 \\
      L18A & 0.40$\pm$0.1 & 0.55 \\
      V19A & 0.32$\pm$0.1 & 0.39 \\
      L33A & 0.27$\pm$0.1 & 0.33 \\
      L36A & 0.25$\pm$0.1 & 0.37 \\
      V37A & 0.15$\pm$0.1 & 0.24 \\
      L52A & 0.03$\pm$0.1 & 0.00 \\
      V68A & 0.23$\pm$0.1 & 0.10 \\
      V71A & 0.36$\pm$0.1 & 0.07 \\
      A76G & 0.37$\pm$0.2 & 0.05 \\
      A77G & 0.37$\pm$0.1 & 0.05 \\
      F83A & 0.31$\pm$0.1 & 0.52 \\
      \rule{0pt}{6ex}
      \textbf{1tiu} & $\phi_{\text{expt}}$ & $\phi_{\text{pred}}$ \\
      \hline
      I2A & 0.45$\pm$0.1 & 0.00 \\
      V4A & 0.29$\pm$0.1 & 0.00 \\
      L8A & 0.28$\pm$0.1 & 0.03 \\
      V13A & 0.00$\pm$0.1 & 0.01 \\
      V15A & 0.01$\pm$0.1 & 0.01 \\
      A19G & 0.38$\pm$0.1 & 0.25 \\
      I23A & 0.82$\pm$0.1 & 0.15 \\
      L25A & 0.42$\pm$0.1 & 0.00
    \end{tabular}
  \end{minipage}
  \caption{\textbf{List of $\phi$-value mutations.}  Data points are from the following references, modified as described in \secref{sec:phi_psi}: 1enh~\cite{gianni2003unifying}, 1igd~\cite{mccallister2000critical}, 1shg~\cite{martinez1999folding}, 1k53~\cite{kim2000breakdown}, 2ci2~\cite{itzhaki1995structure}, 1csp~\cite{garcia2004folding}, 1ubq~\cite{went2005ubiquitin} refolding, 1imp~\cite{friel2003structural} and 1tiu~\cite{fowler2001mapping}.}
  \label{tab:phi1}
\end{table*}
\begin{table*}
  \begin{minipage}[t]{4.25cm}
    \begin{tabular}{lll}
      V30A & 0.45$\pm$0.1 & 0.00 \\
      G32A & 0.51$\pm$0.1 & 0.74 \\
      L36A & 0.50$\pm$0.1 & 0.53 \\
      L41A & 0.40$\pm$0.1 & 0.00 \\
      C47A & 0.42$\pm$0.1 & 0.48 \\
      H56A & 0.52$\pm$0.1 & 0.51 \\
      L58A & 0.79$\pm$0.1 & 0.65 \\
      L60A & 0.67$\pm$0.1 & 0.16 \\
      C63A & 0.23$\pm$0.1 & 0.04 \\
      M67A & 0.13$\pm$0.1 & 0.11 \\
      V71A & 0.63$\pm$0.1 & 0.68 \\
      A82G & 0.16$\pm$0.1 & 0.30 \\
      L84A & 0.05$\pm$0.1 & 0.31 \\
      V86A & 0.01$\pm$0.1 & 0.02 \\
      \rule{0pt}{6ex}
      \textbf{1btb} & $\phi_{\text{expt}}$ & $\phi_{\text{pred}}$ \\
      \hline
      Q9G & 0.72$\pm$0.2 & 0.04 \\
      I13A & 0.45$\pm$0.2 & 0.06 \\
      Q18G & 0.69$\pm$0.2 & 0.09 \\
      A25G & 0.68$\pm$0.2 & 0.04 \\
      A36G & 0.70$\pm$0.2 & 0.42 \\
      L37A & 0.59$\pm$0.2 & 0.60 \\
      L41A & 0.45$\pm$0.2 & 0.58 \\
      V45A & 0.47$\pm$0.2 & 0.21 \\
      L49A & 0.47$\pm$0.2 & 0.58 \\
      V50G & 0.77$\pm$0.2 & 0.39
    \end{tabular}
  \end{minipage}
  \begin{minipage}[t]{4.25cm}
    \begin{tabular}{lll}
      F56A & 0.35$\pm$0.2 & 0.13 \\
      Q58G & 0.11$\pm$0.2 & 0.08 \\
      Q61G & 0.09$\pm$0.2 & 0.08 \\
      T63A & 0.38$\pm$0.2 & 0.05 \\
      A67G & 0.30$\pm$0.2 & 0.29 \\
      E68A & 0.52$\pm$0.2 & 0.52 \\
      V70A & 0.41$\pm$0.2 & 0.47 \\
      Q72G & 0.81$\pm$0.2 & 0.85 \\
      A77G & 0.90$\pm$0.2 & 0.73 \\
      A79G & 0.63$\pm$0.2 & 0.81 \\
      T85A & 0.51$\pm$0.2 & 0.64 \\
      \rule{0pt}{6ex}
      \textbf{1fkb} & $\phi_{\text{expt}}$ & $\phi_{\text{pred}}$ \\
      \hline
      V2A & 0.39$\pm$0.1 & 0.39 \\
      V4A & 0.27$\pm$0.1 & 0.40 \\
      T21A & 0.40$\pm$0.1 & 0.45 \\
      V23A & 0.52$\pm$0.1 & 0.47 \\
      V24A & 0.38$\pm$0.1 & 0.45 \\
      T27A & 0.28$\pm$0.1 & 0.41 \\
      F36A & 0.15$\pm$0.1 & 0.29 \\
      L50A & 0.39$\pm$0.1 & 0.33 \\
      V55A & 0.08$\pm$0.1 & 0.32 \\
      I56A & 0.19$\pm$0.1 & 0.34 \\
      R57G & 0.14$\pm$0.1 & 0.37 \\
      E60G & 0.13$\pm$0.1 & 0.26 \\
      E61G & 0.20$\pm$0.1 & 0.36
    \end{tabular}
  \end{minipage}
  \begin{minipage}[t]{4.25cm}
    \begin{tabular}{lll}
      V63A & 0.51$\pm$0.1 & 0.36 \\
      T75A & 0.24$\pm$0.1 & 0.41 \\
      I76A & 0.34$\pm$0.1 & 0.40 \\
      I91A & 0.04$\pm$0.1 & 0.00 \\
      L97A & 0.23$\pm$0.1 & 0.40 \\
      V98A & 0.27$\pm$0.1 & 0.44 \\
      V101A & 0.57$\pm$0.1 & 0.44 \\
      L106A & 0.35$\pm$0.1 & 0.42 \\
      \rule{0pt}{6ex}
      \textbf{1rnb} & $\phi_{\text{expt}}$ & $\phi_{\text{pred}}$ \\
      \hline
      N5A & 0.09$\pm$0.1 & 0.13 \\
      T6G & 0.21$\pm$0.1 & 0.38 \\
      V10A & 0.33$\pm$0.1 & 0.35 \\
      L14A & 0.59$\pm$0.1 & 0.37 \\
      T26G & 0.00$\pm$0.1 & 0.07 \\
      V36A & 0.00$\pm$0.1 & 0.00 \\
      N58A & 0.94$\pm$0.1 & 0.00 \\
      N77A & 0.00$\pm$0.1 & 0.11 \\
      N84A & 0.16$\pm$0.1 & 0.00 \\
      S91A & 0.93$\pm$0.1 & 0.53 \\
      S92A & 0.95$\pm$0.1 & 0.52 \\
      \rule{0pt}{6ex}
      \textbf{3chy} & $\phi_{\text{expt}}$ & $\phi_{\text{pred}}$ \\
      \hline
      A36G & 0.75$\pm$0.1 & 0.66 \\
      D38G & 0.60$\pm$0.1 & 0.33 \\
      A42G & 0.68$\pm$0.1 & 0.62
    \end{tabular}
  \end{minipage}
  \begin{minipage}[t]{4.25cm}
    \begin{tabular}{lll}
      D64A & 0.11$\pm$0.1 & 0.28 \\
      A97G & 0.00$\pm$0.1 & 0.09 \\
      A98G & 0.03$\pm$0.1 & 0.07 \\
      T112G & 0.12$\pm$0.1 & 0.04 \\
      \rule{0pt}{6ex}
      \textbf{2vil} & $\phi_{\text{expt}}$ & $\phi_{\text{pred}}$ \\
      \hline
      L3A & 0.35$\pm$0.1 & 0.00 \\
      V7A & 0.45$\pm$0.1 & 0.00 \\
      I18A & 0.49$\pm$0.1 & 0.44 \\
      I23A & 0.65$\pm$0.1 & 0.62 \\
      M28A & 0.58$\pm$0.1 & 0.65 \\
      C44A & 0.85$\pm$0.1 & 0.64 \\
      V46A & 0.69$\pm$0.1 & 0.66 \\
      L47A & 0.43$\pm$0.1 & 0.58 \\
      L48A & 0.62$\pm$0.1 & 0.67 \\
      I61A & 0.05$\pm$0.1 & 0.71 \\
      L65A & 0.24$\pm$0.1 & 0.61 \\
      E73A & 0.69$\pm$0.1 & 0.56 \\
      A77G & 0.52$\pm$0.1 & 0.57 \\
      A78G & 0.56$\pm$0.1 & 0.58 \\
      T81A & 0.75$\pm$0.1 & 0.57 \\
      M84A & 0.68$\pm$0.1 & 0.57 \\
      L114A & 0.03$\pm$0.1 & 0.01
    \end{tabular}
  \end{minipage}
  \caption{\textbf{List of $\phi$-value mutations (continued).}  Data points are from the following references, modified as described in \secref{sec:phi_psi}: 1btb~\cite{nolting1997folding}, 1fkb~\cite{fulton1999mapping}, 1rnb~\cite{serrano1992folding}, 3chy~\cite{lopez1996structure} and 2vil~\cite{choe2000differential}.}
  \label{tab:phi2}
\end{table*}

\begin{table*}
  \begin{minipage}[t]{5.5cm}
    \begin{tabular}{lll}
      \textbf{1igd} & $\psi_{\text{expt}}$ & $\psi_{\text{pred}}$ \\
      \hline
      K4--T51 & 0.17$\pm$0.1 & 0.55 \\
      I6--T53 & 0.71$\pm$0.1 & 0.51 \\
      N8--T55 & 0.30$\pm$0.1 & 0.26 \\
      T16--Y33 & 0.24$\pm$0.1 & 0.08 \\
      K28--Q32 & 0.24$\pm$0.1 & 0.35 \\
      Q32--D36 & 0.03$\pm$0.1 & 0.33 \\
      T44--T53 & 0.93$\pm$0.1 & 0.61 \\
      D46--T51 & 0.90$\pm$0.1 & 0.66 \\
      \rule{0pt}{6ex}
      \textbf{1k53} & $\psi_{\text{expt}}$ & $\psi_{\text{pred}}$ \\
      \hline
      N9--T19 & 0.75$\pm$0.2 & 0.57 \\
      N9--N59 & 1.00$\pm$0.4 & 0.68 \\
      I11--K61 & 1.00$\pm$0.1 & 0.57 \\
      K28--E32 & 0.26$\pm$0.1 & 0.31 \\
      A35--T39 & 0.00$\pm$0.1 & 0.25 \\
      D50--N59 & 1.00$\pm$0.1 & 0.43 \\
      A52--T57 & 1.00$\pm$0.1 & 0.50
    \end{tabular}
  \end{minipage}
  \begin{minipage}[t]{5.5cm}
    \begin{tabular}{lll}
      \textbf{1ubq} & $\psi_{\text{expt}}$ & $\psi_{\text{pred}}$ \\
      \hline
      Q2--E16 & 0.53$\pm$0.1 & 0.66 \\
      Q2--E64 & 0.03$\pm$0.1 & 0.59 \\
      F4--T12 & 1.00$\pm$0.1 & 0.90 \\
      F4--T66 & 0.75$\pm$0.1 & 0.90 \\
      K6--T12 & 1.00$\pm$0.1 & 0.89 \\
      K6--T66 & 1.00$\pm$0.1 & 0.88 \\
      K6--H68 & 0.52$\pm$0.1 & 0.89 \\
      E24--A28 & 0.48$\pm$0.1 & 0.64 \\
      A28--D32 & 0.90$\pm$0.1 & 0.66 \\
      R42--Q49 & 0.07$\pm$0.1 & 0.40 \\
      R42--H68 & 0.26$\pm$0.1 & 0.72 \\
      R42--V70 & 0.57$\pm$0.1 & 0.67 \\
      F44--Q49 & 0.02$\pm$0.1 & 0.40 \\
      I44--V70 & 1.00$\pm$0.1 & 0.81
    \end{tabular}
  \end{minipage}
  \begin{minipage}[t]{5.5cm}
    \begin{tabular}{lll}
      \textbf{2acy} & $\psi_{\text{expt}}$ & $\psi_{\text{pred}}$ \\
      \hline
      D10--N81 & 0.70$\pm$0.1 & 0.22 \\
      E12--N79 & 1.00$\pm$0.1 & 0.23 \\
      K24--A28 & 0.01$\pm$0.1 & 0.08 \\
      A28--K32 & 0.00$\pm$0.1 & 0.08 \\
      W38--Q50 & 1.00$\pm$0.1 & 0.80 \\
      Q40--V97 & 0.13$\pm$0.1 & 0.07 \\
      S56--H60 & 0.34$\pm$0.1 & 0.35 \\
      R59--E63 & 1.00$\pm$0.1 & 0.42
    \end{tabular}
  \end{minipage}
  \caption{\textbf{List of $\psi$-value mutations.}  For ubiqutin (1ubq), two experimental $\psi$-values (residue pairs 2--16 and 44--70) involve residues that do not form native contacts in the crystal structure.  We calculated theoretical $\psi$-values for the nearest native contacts in our model, replacing these pairs with contacts 1--16 and 44--68, respectively.  Data points are from the following references, modified as described in \secref{sec:phi_psi}: 1igd~\cite{baxa2015even}, 1k53~\cite{yoo2012folding}, 1ubq~\cite{sosnick2004differences} and 2acy~\cite{pandit2006small}.}
  \label{tab:psi}
\end{table*}


\end{document}